\documentclass{article}
\usepackage[utf8]{inputenc}
\usepackage{enumitem}
\usepackage{graphicx}
\usepackage{subcaption}
\usepackage{amssymb}
\usepackage{amsmath}
\usepackage{bm}
\usepackage{geometry}
\usepackage{rotating}
\RequirePackage{geometry}
\usepackage{xcolor}
\usepackage{booktabs}
\usepackage{pdflscape}
\usepackage{longtable}
\usepackage{multirow}
\usepackage[english]{babel}
\usepackage[authoryear,round]{natbib}
\renewcommand{\cite}{\citep}

\usepackage{lineno,hyperref}
\modulolinenumbers[1] % for reviewers

\usepackage{todonotes}
\let\opentask\todo % redefining todo{} as todo[inline]{} to appear in the text not in margins
\renewcommand{\todo}[1]{\opentask[inline,color=red!40]{#1}}

\title{Can Human Drivers and Connected Autonomous Vehicles Co-exist in Lane-Free Traffic? A Microscopic Simulation Perspective}
\author{Arslan Ali Syed\thanks{Corresponding author}, Majid Rostami-Shahrbabaki, and Klaus Bogenberger\thanks{arslan.syed@tum.de; majid.rostami@tum.de; klaus.bogenberger@tum.de; All authors are with the Chair of Traffic Engineering and Control, Technical University of Munich, Germany.}}
\date{\today}

\begin{document}

\maketitle

\section*{Abstract}

Recent advancements in connected autonomous vehicle (CAV) technology have sparked growing research interest in lane-free traffic (LFT). LFT envisions a scenario where all vehicles are CAVs, coordinating their movements without lanes to achieve smoother traffic flow and higher road capacity. This potentially reduces congestion without building new infrastructure. However, the transition phase will likely involve non-connected actors such as human-driven vehicles (HDVs) or independent AVs sharing the roads. This raises the question of how LFT performance is impacted when not all vehicles are CAVs, as these non-connected vehicles may prioritize their own benefits over system-wide improvements.
This paper addresses this question through microscopic simulation on a ring road, where CAVs follow the potential lines (PL) controller for LFT, while HDVs adhere to a strip-based car-following model. The PL controller is also modified for safe velocities to prevent collisions. The results reveal that even a small percentage of HDVs can significantly disrupt LFT flow: 5\% HDVs can reduce LFT’s maximum road capacity by 20\% and a 40\% HDVs nearly halves it, up until 100\% HDVs where it drops by nearly 60\%. The study also develops an adaptive potential line (APL) controller that forms APL corridors in the surroundings of HDVs. APL shows a peak traffic flow improvement of nearly 10\% over the PL controller. The study indicates that a penetration rate of approximately 60\% CAVs is required to start observing the major LFT benefits. These findings open a new research direction on minimizing the adverse effects of non-connected vehicles on LFT.

\section*{Comments on the current draft}
This version corresponds to the final published article in Transportation Research Part C: Emerging Technologies (DOI: \url{https://doi.org/10.1016/j.trc.2025.105315}). It incorporates revisions made during peer review, including an improved literature review, clearer methodological descriptions, and explicit consideration of safety and comfort. These changes affected some simulation results (e.g., maximum road capacity in the all-CAV scenario, impacts of HDV penetration, and improvements via the Adaptive Potential Lines method). However, the main conclusions remain unchanged—for instance, a CAV penetration of 60\% is still required before major advantages of LFT are observed. The above-linked article is published under the Creative Commons Attribution 4.0 International (CC BY 4.0) license (\url{https://creativecommons.org/licenses/by/4.0/}).

\section*{Keywords}
Lane-Free Traffic, Lane-less Traffic, Human Driver Model, Mixed Traffic, Connected Autonomous Vehicles (CAVs), Ring Road, Adaptive Potential Line

\section*{Highlights}

% You are required to provide article highlights at submission.

% Highlights are a short collection of bullet points that should capture the novel results of your research as well as any new methods used during your study. Highlights will help increase the discoverability of your article via search engines. Some guidelines:

% Submit highlights as a separate editable file in the online submission system with the word "highlights" included in the file name.

% Highlights should consist of 3 to 5 bullet points, each a maximum of 85 characters, including spaces.

\begin{itemize}
    \item Even a small percentage of human driver vehicles (HDVs) significantly lowers the LFT flow.
    \item The LFT traffic flow drops by 20\% with just 5\% HDV penetration rate
    \item 60\% CAV Penetration rate is required before observing LFT’s major advantages
    \item Introduced Adaptive Potential Line (APL) controller  to improve the LFT with HDVs
    \item A peak average improvement of nearly 10\% was observed via the APL strategy over PL controller
\end{itemize}

\section{Introduction}

With urbanization and increased usage of private vehicles (PVs), traffic congestion has been an ever-increasing problem, especially in cities. The increased traffic has led to worldwide road construction projects, yet the problem persists \cite{rahmanTrafficCongestionIts2022}. The last two decades have seen significant changes in the automotive industry. There has been a steady increase in autonomous vehicle (AV) technology. Multitudes of works have suggested using shared AVs (SAVs) to sway people away from using PVs in the direction of public transport (PT) by providing first- or last-mile operation via SAVs or providing a whole mobility-on-demand (MOD) service using AV fleet \cite{narayananSharedAutonomousVehicle2020, syedUserAssignmentStrategyConsidering2021,syedNeuralNetworkBased2019}. However, the experience with the current MOD services indicates that if not appropriately regulated, these services may contribute further to the traffic congestion with additional vehicles \cite{rahmanTrafficCongestionIts2022}. Therefore, the fundamental problem remains that a vehicle occupies the same amount of road space regardless of whether a human or AV technology drives the vehicle and would require increased road infrastructure to accommodate more vehicles.

\begin{figure}[tb]
     \centering
     \begin{subfigure}[b]{.32\textwidth}
         \centering
         \includegraphics[height=2cm]{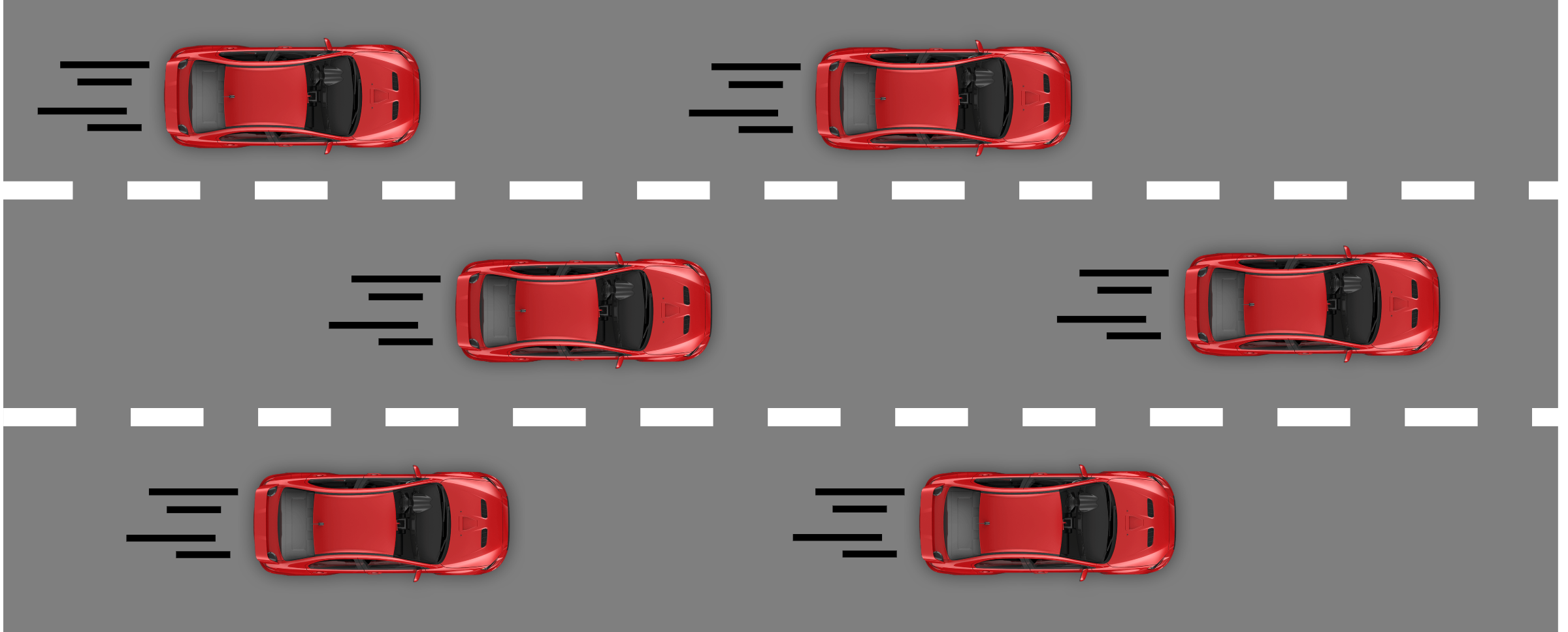}
         \caption{Lane-based traffic}
         \label{fig: Lane based vs Lane Free Traffic (a)}
     \end{subfigure}
     \hfill
     \begin{subfigure}[b]{.32\textwidth}
         \centering
         \includegraphics[height=2cm]{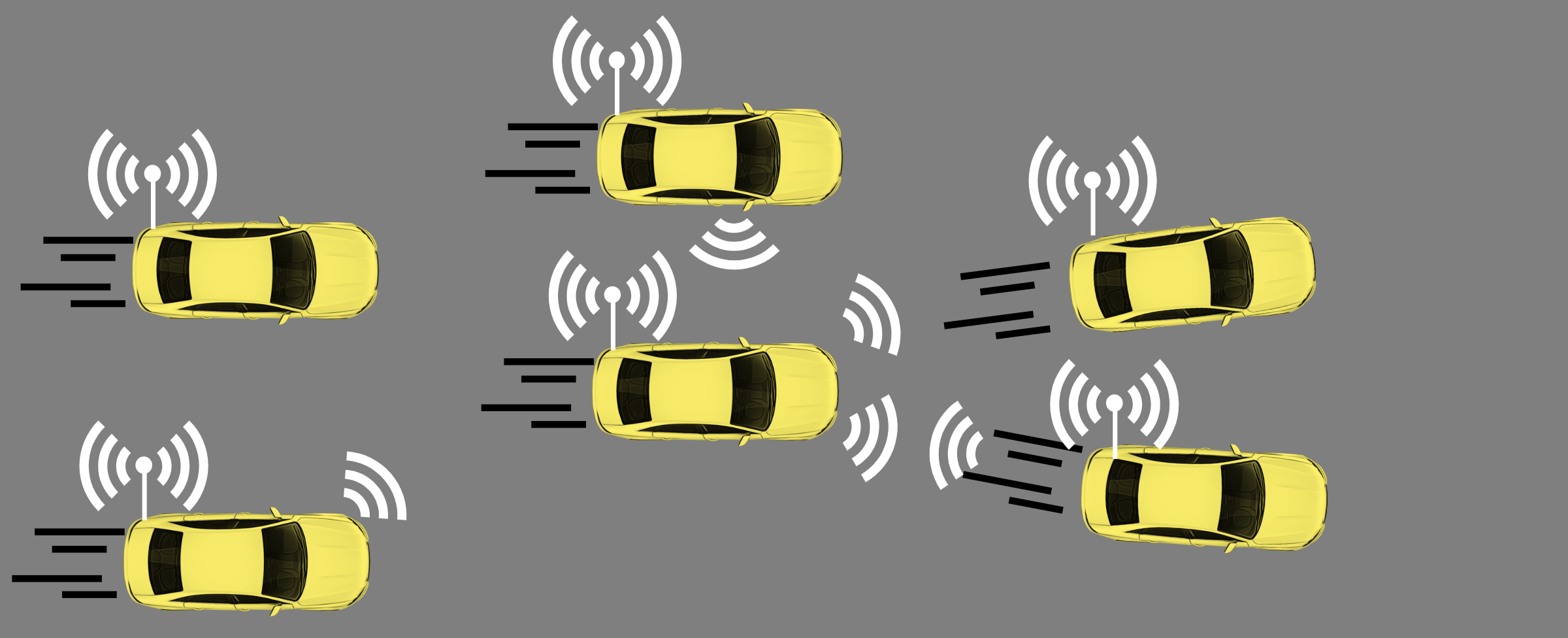}
         \caption{LFT with CAVs}
         \label{fig: Lane based vs Lane Free Traffic (b)}
     \end{subfigure}
     \hfill
     \begin{subfigure}[b]{0.32\textwidth}
         \centering
         \includegraphics[height=2cm]{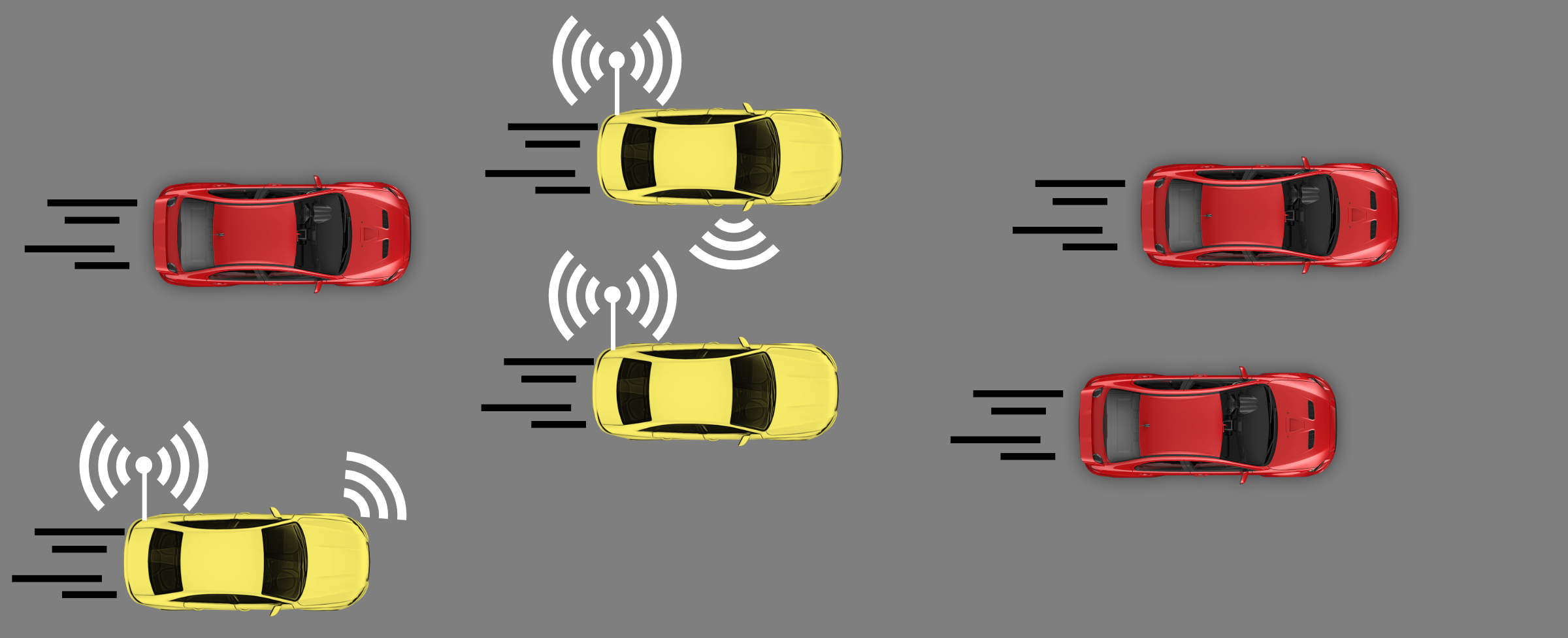}
         \caption{LFT with CAVs and HDVs}
         \label{fig: Lane based vs Lane Free Traffic (c)}
     \end{subfigure}
        \caption{Traditional lane-based traffic, lane-free traffic (LFT) and LFT wtih a mixture of CAVs and HDVs.}
        \label{fig: Lane based vs Lane Free Traffic}
\end{figure}

To address the above problem, there has also been research on increasing road capacities and safety by utilizing new AV technologies. In this regard, connected autonomous vehicles (CAVs) play a significant role \cite{narayananFactorsAffectingTraffic2020a, ahmed2022technology}. The real-time data sharing from vehicle-to-vehicle (V2V), vehicle-to-infrastructure (V2I), or vehicle-to-everything (V2X) allows the CAVs to constantly observe the environment from multiple perspectives, enabling technologies such as advanced traffic state estimation approaches \cite{majid-state}, cooperative adaptive cruise control (CACC) \cite{ahmed2022technology}, speed advisory systems \cite{majid-speed} or integrated intersection control \cite{nielsSimulationBasedEvaluationNew2020}. The most significant change that CAVs can bring is perhaps in the currently dominant driving paradigm of lane-based traffic management --- the CAVs show the potential to drive even when no specific lanes are marked on the road \cite{papageorgiou2021lane}. The concept of managed lanes was introduced in the last century to help coordinate the vehicle movements by human drivers in an era of ever-increasing maximum possible vehicle speed \cite{MayaSOA}. With the introduction of CAVs, the fixed lanes can be removed since the CAVs can coordinate their movements via communication channels. This has led to the concept of lane-free traffic (LFT) for CAVs \cite{papageorgiou2021lane}. Figure~\ref{fig: Lane based vs Lane Free Traffic} shows the concept of LFT. As shown in Figure~\ref{fig: Lane based vs Lane Free Traffic (b)}, the LFT vehicles can communicate and influence the movements of upstream and downstream vehicles using wireless communication. Usually, the LFT controller algorithms achieve this by assuming artificial nudging and repulsive forces applied to vehicles in the front and the back, respectively \cite{papageorgiou2021lane, yanumulaOptimalPathPlanning2021, rostami2023increasing}.

The LFT was inspired by the vehicles driving without strict lane discipline in some countries, referred to as lane-less traffic \cite{MayaSOA,papageorgiou2021lane}. A key motivating factor was the significantly greater utilization of road width observed in lane-less traffic, as opposed to the roughly 50\% occupancy typical in lane-based systems. This 50\% estimate is based on a comparison between the average widths of vehicles (cars and trucks) and standard lane widths\cite{papageorgiou2021lane}. While lane-less traffic enables higher road occupancy, it also introduces a substantially increased risk of accidents \cite{mukherjee2022development}
. However, with the automation of vehicles in LFT, such risks can be mitigated through the integration of advanced safety features within the control algorithms. Apart from fast and accurate decisions in LFT, another major difference between lane-less and LFT traffic is that while in the former, communication with other drivers is limited to honking, headlight flashing, or hand gesturing, the vehicles in the latter can fully communicate their intended trajectories and other important information to significantly larger areas. This allows LFT vehicles to coordinate their movements in a way that substantially improves the system-wide maximum flow rate compared to lane-based traffic for the same road width \cite{rostami2023increasing}. However, the exact flow rate improvement depends on the LFT controller used. %It ranges from a rate of X $veh/h$ for a density of Y $veh/km$ \greentext{[cite]} to as high as X $veh/h$ for the same vehicle density \greentext{[cite]}.

While the LFT strategy has a high potential to improve traffic conditions, its realization may still be decades away. First, significant technological developments are required to ensure flawless inter-vehicular communication and reliable autonomous driving functionality. Second, the CAVs in LFT will be driving in a completely new traffic mode where safety will be of primary concern. Consequently, LFT must undergo rigorous testing before any CAV can be driven in LFT mode. However, unlike the functionality of current AVs, which can be tested with a single vehicle, the LFT would require tests with multiple CAVs operating in an unexplored driving mode. To solve this problem, some researchers have suggested using a driving simulator to study the safety as perceived by the traveler in LFT \cite{sekeranInvestigatingLaneFreeTraffic2023}. Nonetheless, it is expected that even if the cities are convinced of investing in the LFT, the transition phase or even the actual operation of LFT may still involve some traffic participants who do not necessarily coordinate their movements with other vehicles. This can happen for multiple reasons, for example, the involvement of human drivers, having an older AV without a communication module, or a CAV temporarily losing its communication capability. Under all of these circumstances, the question remains: how will the efficiency of the LFT be affected by it? It is equally important to ask how much the proportion of CAVs should be to achieve the benefits of LFT.

This work attempts to answer the above questions. The paper mainly focuses on the impacts of having some traffic participants in LFT who try to improve self-interest instead of coordinating and optimizing the flow of the whole system. These participants are assumed to be human drivers trying to achieve their desired speeds while maintaining a safe distance from the vehicle in the front. The LFT traffic vehicles are assumed to be unable to exert nudging forces on these HDVs, as shown in Figure~\ref{fig: Lane based vs Lane Free Traffic (c)}, which is expected to disrupt the functioning of the LFT controller. The paper does not explicitly model independent AV; instead, it is assumed that these HDVs can also partially represent independent AVs since AVs may already be trying to replicate human driving behavior. Nonetheless, the main focus of the paper is not on accurately modeling the human or independent AV's behavior but rather on the disruption it causes to the LFT.

\begin{figure}[tb]
    \centering
    \includegraphics[width=1\linewidth]{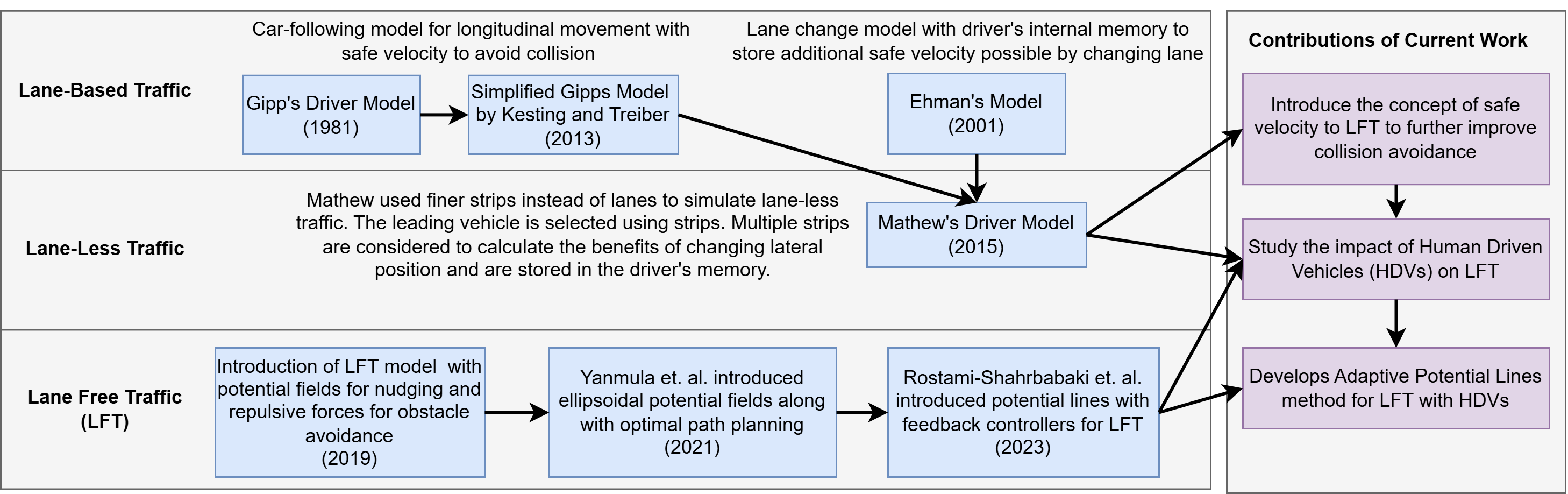}
    \caption{The main contributions of the current work and its relation to the available literature.}
    \label{fig: contrbutions}
\end{figure}

To study the impact of the whole setup, microscopic simulations of a 1~$km$ ring road are used. The simulation builds upon prior research, extending and refining earlier contributions. Figure~\ref{fig: contrbutions} illustrates these foundational works and their relationship to the key contributions of the present study. The human drivers are modeled using the car-following model by \citet{mathewStripBasedApproachSimulation2015}, designed to simulate lane-less traffic. For the CAVs, the paper uses the potential line (PL) controller introduced by \citet{rostami2023increasing}. Inspired by the human model of \cite{mathewStripBasedApproachSimulation2015}, the paper also introduces the concept of safe acceleration into the PL controller as a secondary measure to avoid any potential collision. Finally, an adaptive potential lines (APL) controller is introduced that improves the flow of CAVs mixed with HDVs in LFT. The new controller forms APL corridors with modified PLs in the vicinity of HDVs.

\section{Background}
\label{sec: background}

To address the ever-increasing traffic demand, modern roads were designed to accommodate more vehicles and alleviate congestion by rerouting traffic away from city centers. This includes the introduction of beltways or ring roads, which provide more direct and faster connections to areas around the city. However, investigations worldwide reveal the immediate benefits of beltways have been diminished due to the induced demand and relocation of jobs and housing to suburban areas \cite{ring1,ring2,ring3}. Consequently, innovative approaches using advanced Vehicle Automation and Communication Systems (VACS) are gaining the center stage for sustainable long-term solutions to ensure smooth traffic flow, reduce congestion, and enhance safety \cite{diakaki2015overview}. In the last decade, the application of CAVs comprised the vast majority of cutting-edge research towards addressing traffic congestion \cite{ahmed2022technology}, including LFT for more exploitation of the lateral capacity of the roads \cite{MayaSOA}. 

\begin{figure}[tb]
     \centering
     \begin{subfigure}[b]{.49\textwidth}
         \centering
         \includegraphics[height=2.9cm]{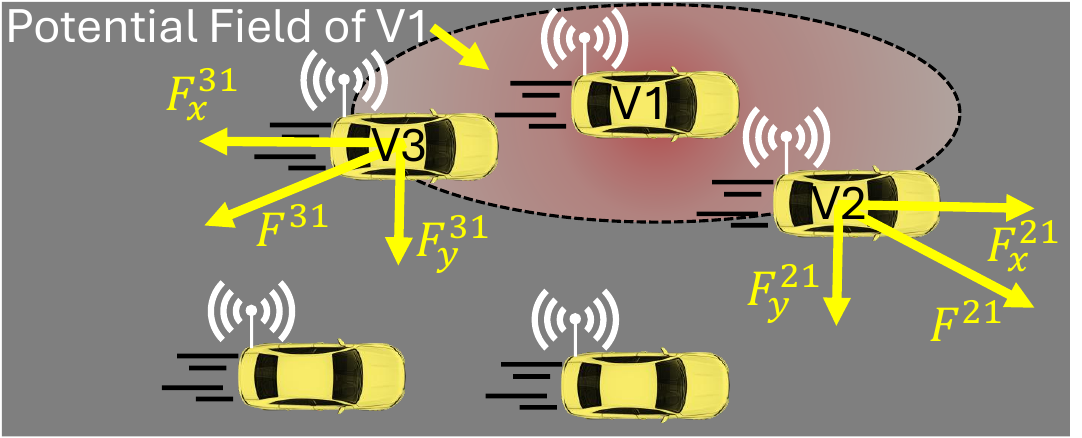}
         \caption{Potential field and artificial forces in LFT}
         \label{fig: Lane Free Traffic Controller (a)}
     \end{subfigure}
     \hfill
     \begin{subfigure}[b]{.49\textwidth}
         \centering
         \includegraphics[height=3.3cm]{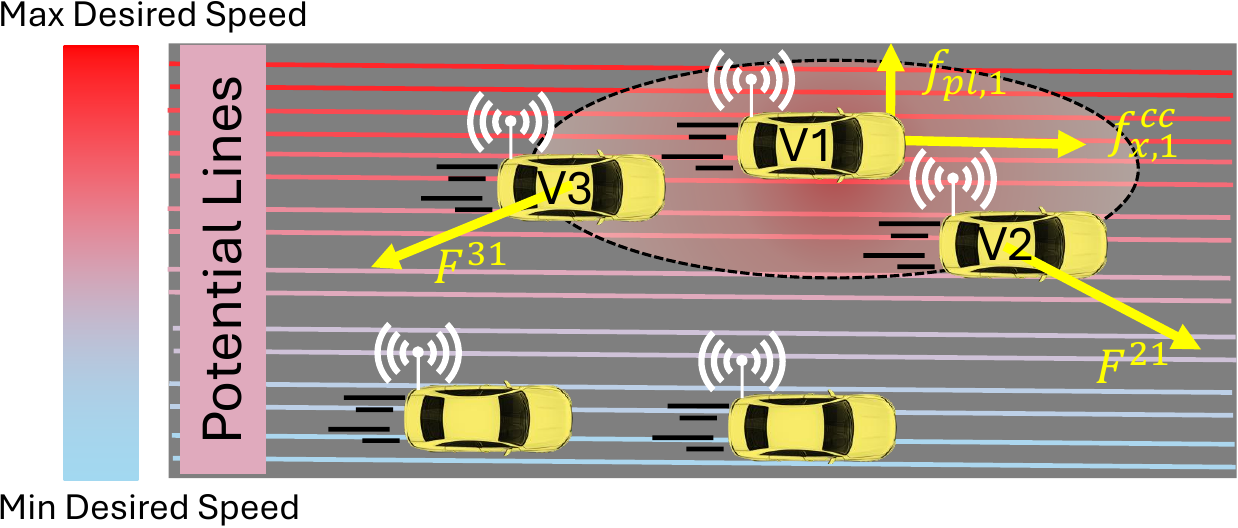}
         \caption{PL Controller}
         \label{fig: Lane Free Traffic Controller (b)}
     \end{subfigure}
     \hfill
     \begin{subfigure}[b]{0.49\textwidth}
         \centering
         \includegraphics[height=3.3cm]{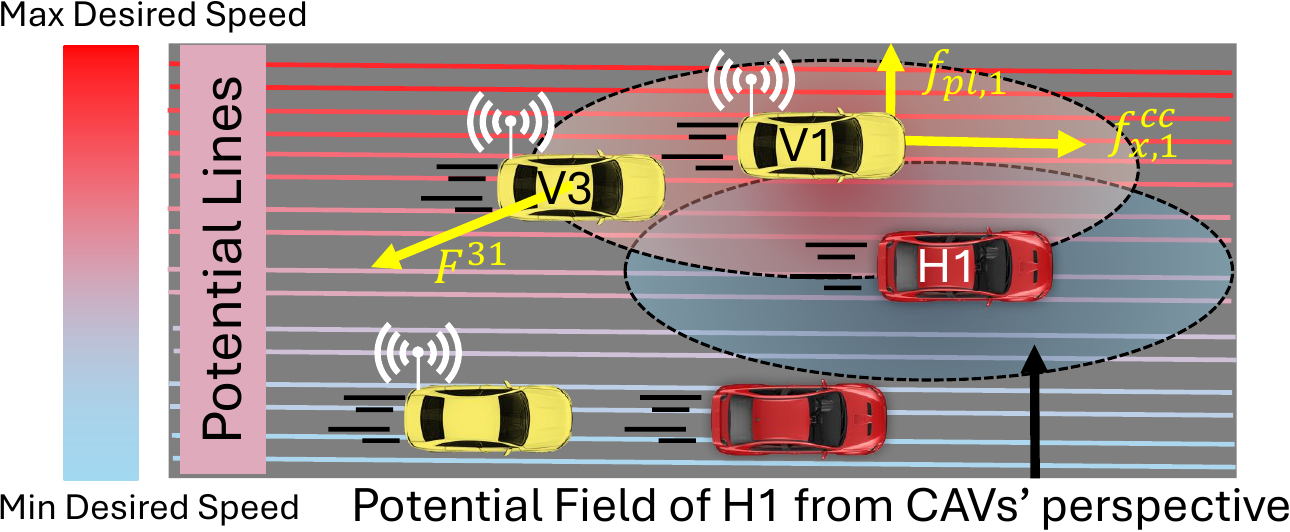}
         \caption{PL Controller with a mixture of CAVs and HDVs}
         \label{fig: Lane Free Traffic Controller (c)}
     \end{subfigure}
        \caption{Potential fields, artificial forces and potential line (PL) controller for LFT.}
        \label{fig: Lane Free Traffic Controller}
\end{figure}

Since LFT allows for lateral freedom of CAVs, compared to lane-based traffic, novel driving strategies have been proposed for vehicle navigation in this new environment \cite{Papageorgiou2024}. A common approach comprises defining artificial potential fields around each vehicle for collision avoidance \cite{malekzadeh2022empirical}, as shown in Figure~\ref{fig: Lane Free Traffic Controller (a)}, and including additional controllers for achieving other objectives such as driving close to the desired speed, staying within the road boundary, and driving energy-efficiently. \citet{Karteek} formulated all the objectives in an optimal control problem and solved it for each vehicle in real-time. A more structured lane-free traffic is proposed in \cite{potentialline}. In this approach, an artificial potential line (PL) is assigned for each vehicle as the desired lateral location, as illustrated in \ref{fig: Lane Free Traffic Controller (b)}. This led to more laminar traffic flow, eliminating unnecessary lateral movement of vehicles. Other approaches, such as nonlinear controllers \cite{cruise-ringroad}, have also been proposed that use a more complex dynamic model of vehicles. While the majority of LFT research focuses on freeway networks, few have considered elements of urban networks such as beltways \cite{rostami2023increasing}, intersections \cite{StugerLFInt,MalcolmLFInt}, and roundabouts \cite{nader}. In addition to conventional control approaches, reinforcement learning-based approaches have also been used, showcasing comparable benefits for driving comfort and traffic efficiency \cite{berahman-ma,BERAHMAN-ddpg,DRL_Karalakou,DDPG_Karalakou}. In addition, LFT calls for novel ideas such as vehicle flocking \cite{majid-flockingone,majid-flocktwo} or snake-like platooning \cite{DABESTANI202461}. Table~\ref{table: literature review} presents an overview of the LFT control methods.

It should be noted that almost all current LFT driving strategies rely highly on V2V communication, sharing the vehicle's current state or even planned trajectory. For example, to apply an optimal control approach, the vehicles in \cite{Karteek} share planned trajectories with the surrounding vehicles. Since the HDVs in the current work are assumed to make independent decisions by considering self-interests, their trajectories are not available to the LFT controller and, thus, cannot be shared with CAVs. Therefore, these latter types of LFT controllers pose a significant challenge in handling HDVs.

To simulate mixed HDVs and CAVs traffic, it is essential to have a driving model that accurately represents human drivers. Currently, there is no data on how humans would drive in an LFT environment. Therefore, the most suitable approach is to consider models developed for lane-less traffic. In the literature, models for lane-less traffic often target a traffic condition where vehicles of various types and sizes share the same road without strict lane discipline. This can include vehicles like cars, rickshaws, motorbikes, and trucks \cite{munigetyBehavioralModelingDrivers2016, ravishankarVehicleTypeDependentCarFollowing2011,martin-particle}.

The main challenge in modeling lane-less traffic is that most car-following models assume homogeneous traffic driving in lanes. These models typically assume a pair of a leader and a follower vehicle, which suits lane-based traffic and focuses mainly on longitudinal movements. However, lane-less traffic has certain additional characteristics that are not found in lane-based traffic. For example, smaller vehicles tend to squeeze into gaps between larger vehicles, exhibit staggered following, and show lateral shifting or even multiple leaders \cite{munigetyBehavioralModelingDrivers2016,AnalysisVehicleFollowingBehavior2020}.

The literature shows a gradual improvement in models that can deal with lane-less traffic. For instance, \citet{gunayCarFollowingTheory2007} incorporated non-lane-based car following into the basic model of \citet{gippsBehaviouralCarfollowingModel1981}. Due to the off-centered vehicle positions, this model does not allow full leadership to the front vehicle, and thus making staggered following possible. Although the model is designed for homogeneous traffic, it allows weak lane discipline behavior and includes lateral discomfort. \citet{ravishankarVehicleTypeDependentCarFollowing2011} modified \cite{gippsBehaviouralCarfollowingModel1981} to include type-dependent variables for the leader-follower pair. Later, inspired by approaches with a discretized lateral axis \cite{georgeoketchNewModelingApproach2000}, they improved the type-dependent model using the concept of strips \cite{mathewStripBasedApproachSimulation2015}. A similar concept of sublane was later introduced into Simulation of Urban Mobility (SUMO) \cite{behrisch2011sumo} for Chinese traffic situations and became part of its standard distribution \cite{semrau2016simulation}. Other methods have also been used for mixed traffic \cite{hossainModellingTrafficOperations1998,arasanMethodologyModelingHighly2005}. More recently, with the possibility of collecting lane-less traffic data more efficiently, data-driven approaches have emerged in the literature \cite{AnalysisVehicleFollowingBehavior2020,papathanasopoulou2018flexible}. However, strip-based approaches have so far been simple and computationally effective for simulating lane-less traffic.

\begin{landscape}
\footnotesize

\begin{longtable}{@{}lllp{14cm}@{}}
\toprule
\textbf{Study} &
  % \textbf{\begin{tabular}[c]{@{}l@{}}LFT Controller\\ Type\end{tabular}} &
  \textbf{\begin{tabular}[c]{@{}l@{}}Environment \\ Setting\end{tabular}} &
  \textbf{\begin{tabular}[c]{@{}l@{}}Traffic \\ Participants\end{tabular}} &
  \textbf{LFT Control Method} \\ \midrule  \endfirsthead 
\citet{papageorgiou2021lane} &
  % Decentralized &
  Ring Road &
  CAVs &
  A conceptual work with target speed and potential field-based LFT control. According to the detailed version of the work \cite{papageorgiou2019lane}, the target speed controller uses Guass error function for cruise control and the potential field is based on $\Pi$ function. \\
\citet{yanumulaOptimalPathPlanning2021} &
  % Centralized &
  Ring Road &
  CAVs &
  The work introduces an ellipsoidal potential field function combined. The LFT is formulated as a non-linear optimal control problem with multiple objectives and a model predictive control approach. \\
\citet{malekzadeh2022empirical} &
  % Decentralized &
  Ring Road &
  CAVs &
  Extends the controller of \citet{papageorgiou2021lane}. The forces are based on adaptive cruise control (ACC)-inspired logic using an Action Graph for collision avoidance and a boundary feedback controller to keep vehicles within the road edges. \\
\citet{BERAHMAN-ddpg} &
  % Decentralized &
  Ring Road &
  CAVs &
  Combines the ellipsoidal artificial potential field with deep reinforcement learning (RL) based LFT control. \\
\citet{DRL_Karalakou} &
  % Decentralized &
  Ring Road &
  CAVs &
  Test different reward functions for training an RL-based LFT control. \\
\citet{majid-flocktwo} &
   % & Decentralized
  Ring Road &
  CAVs & 
  A two-level controller for vehicle flocking is developed, where in the tactical layer, the control mode is defined, and in the operational layer, the individual and group movement of vehicles is controlled.
   \\
\citet{Karteek} &
  % Centralized &
  Ring Road &
  CAVs &
  It builds on top of \citet{yanumulaOptimalPathPlanning2021} by introducing vehicle nudging, improved obstacle avoidance, and adaptive desired speeds, enabling more efficient maneuvers in dense traffic. It also enhances the optimal control formulation with additional objectives for comfort and emergency collision handling. \\
\citet{rostami2023increasing} &
  % Centralized &
  Ring Road &
  CAVs &
  Introduces the concept of assigning specific lateral positions, called potential lines, to vehicles according to their desired speeds. A Proportional controller is then used for lateral control of vehicles. \\
\citet{potentialline} &
  % Centralized &
  \begin{tabular}[c]{@{}l@{}}Motorway \\ with ramps\end{tabular} &
  \begin{tabular}[c]{@{}l@{}}CAVs,\\ Emergency \\ Vehicle\end{tabular} &
  The LFT control is the same as in \cite{rostami2023increasing}. However, it uses the Probability Integral Transform for uniform lateral distribution of potential lines. An adaptation of the potential line strategy is used for emergency vehicle preemption.\\
\citet{cruise-ringroad} &
  % Decentralized &
  \begin{tabular}[c]{@{}l@{}}Ring Road, \\ Straight Road\end{tabular} &
  CAVs &
  It introduces a Lyapunov Function-based Cruise Controller based on a nonlinear bicycle dynamic model for vehicles. \\
\citet{StugerLFInt} &
  % Centralized &
  Intersection &
  CAVs &
  Intersection control using artificial potential fields and PID controller. The LFT control is divided over multiple modules such as: (1) module for flocking vehicles with similar routes (2) module for dynamic entry and exist points (3) module for reserving tiles for traversing the main intersection area. \\
\citet{DDPG_Karalakou} &
  % Decentralized &
  Ring Road &
  CAVs &
  Extends the RL-based LFT controller of \citet{DRL_Karalakou} using   multiple RL algorithms and improves the reward functions used for RL. \\
\citet{majid-flockingone} &
   % & Decentralized
  Ring Road &
  CAVs &
    It proposed an algorithm for vehicular flocking based on (1) a Mexican hat function for flock formation, (2) a consensus algorithm for velocity matching, and (3) a navigation feedback. \\
\citet{nader} &
  % Decentralized &
  \begin{tabular}[c]{@{}l@{}}Large \\ Roundabouts\end{tabular} &
  CAVs &
  Uses OD-specific movement corridors and space-dependent desired orientations to guide each vehicle. Non-linear feedback controllers are used   for circular and straight movement, and linear feedback controllers are used for boundary control. \\
\citet{berahman-ma} &
  % Decentralized &
  \begin{tabular}[c]{@{}l@{}}Ring Road, \\ Motorway with \\ ramps\end{tabular} &
  CAVs &
  Improves the control approach of \citet{BERAHMAN-ddpg} using a multi-agent and multi-task RL approach. The RL agents learn from shared experiences to achieve more   complex driving tasks. \\
\citet{DABESTANI202461} &
  % Centralized &
  Straight Road &
  CAVs &
  It introduces a joint trajectory-based optimization approach forming 1-D snake-like or   2D deformable flocks. It minimizes several objectives like collision avoidance, deviation from desired speed, formation maintenance, etc. \\
\citet{MalcolmLFInt} &
  % Centralized &
  Intersection &
  CAVs, VRUs &
  Uses a novel first come, first served reservation and trajectory planning approach called FERSTT (Fast Exploring Rule-based Spatiotemporal Trajectory Tree) for intersection control with vulnerable road users (VRUs). VRUs are given separate signal phases for   crossing the intersection. \\ \bottomrule

  \caption{Overview of the literature related to LFT control methods.}
  \label{table: literature review}
\end{longtable}

\end{landscape}

\section{Methodology}
\label{section: methodology}

To simulate LFT with CAVs and HDVs, the paper uses two methods simultaneously: a car-following model for HDVs and LFT controller for CAVs. After establishing the equations for vehicle dynamics, the section presents both in detail.

\subsection{Vehicle Dynamics}

The vehicles in the simulation move in discrete time steps using a double integrator model. The model is implemented using differential equations. Let $k$, $\Delta T$ and $t = k\cdot \Delta T$ represent the current time step, time step size and current simulation time, respectively. For a vehicle $i$, let $x_i$, $v_{x,i}$ and $a_{x,i}$ represent the longitudinal position, speed, and acceleration, respectively. Similarly, let $y_i$, $v_{y,i}$, and $a_{y,i}$ represent the same variables for the lateral axis, respectively. Then the equations for vehicle dynamics are given as:
\begin{align}
x_i(k + 1) &= x_i(k) + \Delta T v_{x,i}(k) + \frac{1}{2} \Delta T^2 a_{x,i}(k) \tag{1a} \\
y_i(k + 1) &= y_i(k) + \Delta T v_{y,i}(k) + \frac{1}{2} \Delta T^2 a_{y,i}(k) \tag{1b} \\
v_{x,i}(k + 1) &= v_{x,i}(k) + \Delta T a_{x,i}(k) \tag{1c} \\
v_{y,i}(k + 1) &= v_{y,i}(k) + \Delta T a_{y,i}(k) \tag{1d}
\end{align}

According to the vehicle type (HDV or CAV), each time step calculates the accelerations in both directions using the human driver model or the LFT controller. Then, the above equations are used to calculate the vehicle states for the next time step. The current study also assumes that each vehicle $i$ has a desired longitudinal speed $v_{des,i}$, which it aims to achieve during the simulation. To prevent the vehicle from leaving the road boundary, the simulation models limit the lateral acceleration such that the road boundary is not crossed.

\subsection{Human Driver Model}

The human driver model used is based on the strip-based model proposed by \citet{mathewStripBasedApproachSimulation2015} for lane-less traffic. Besides being simple, the main reason for choosing the model is that it was validated  in a SUMO simulation using video data from a highway in Mumbai, India. This model differs from traditional lane-based simulations in that it allows continuous movement in the lateral axis by discretizing it into strips. The current paper replicates Mathew's model without significant changes. The following summarizes its formulation for the longitudinal and lateral movements.

\subsubsection{Longitudinal Movement}

Mathew's model based the longitudinal movement on the Gipps model \cite{gippsBehaviouralCarfollowingModel1981} originally meant for lane-based traffic. Instead of using the original Gipp's formulation, they used its simplification as presented in \cite{kesting2013traffic}. Nonetheless, both the original and its simplification are conceptually identical.

For a collision-free movement, the model uses a safe velocity $v^{safe}_{x, i}$ with regards to the vehicle in the front, called the leader. This causes the subject vehicle $i$ to maintain a safe gap and stop behind the leading vehicle without collision if required. The formulation for the safe velocity is given as:

\begin{equation}
\label{eq: safe velocity}
    v^{safe}_{x, i}(k+1) = -\tau \cdot A^-_x + \sqrt{(\tau \cdot A^-_x)^2 + v_{x, leader}(k)^2 + 2 \cdot A^-_x \cdot (g_{leader}(k) - g_o)}
\end{equation}
where $\tau$ is the reaction time in seconds ($s$), $g_{leader}$ is the distance from the front of the $i$ to the back of the leader in meters ($m$), $A^-_x$ is the maximum deceleration ability of $i$ in $m/s^2$ and $v_{x, leader}$ is the longitudinal velocity of the leader in $m/s$. $g_o$ is the minimum distance $i$ withholds if the leader suddenly stops.

Eq.~\ref{eq: safe velocity} is the simplified Gipp's model. \citet{mathewStripBasedApproachSimulation2015} extended it to account for different combinations of follower-leader vehicle types (e.g. trucks, rickshaws, motorbikes, cars). More specifically, \citet{mathewStripBasedApproachSimulation2015} replaced the term $(g_{leader} - g_o)$ in Eq.~\ref{eq: safe velocity} by a regression-based formulation to convert different combinations of vehicle types into equivalent gap for car following car. However, since all vehicles in the current study are supposed to be cars, this modification is ignored in the current work. Additionally, $g_0$ is also set to 0 for simplification.

The next important factor in Eq.~\ref{eq: safe velocity} is the determination of the leader vehicle. In lane-based traffic, the leader is the immediate vehicle in the front in the same lane. However, selecting a leader in traffic without lanes is more challenging since multiple vehicles could be in front. \citet{mathewStripBasedApproachSimulation2015} suggested discretizing the lateral axis strips of width $\Delta s$. Instead of occupying a lane, each vehicle occupies a strip. For the vehicle $i$, the leader is chosen by looping through all the vehicles that occupy any of the strips that $i$ occupies and then selecting the one closest to $i$ in the longitudinal distance. For computational efficiency, the current study only considers the vehicles within the front distance $\Delta D_{front}$ of $i$. If no such leading vehicle exists within $\Delta D_{front}$, the $v^{safe}_{x, i}$ is set to the desired speed $v_{des,i}$ of the vehicle.

After determining the $v^{safe}_{x, i}$, the difference to the current speed $v^{diff}_i = v_{safe} - v_{x,i}$ is calculated, which is then used to calculate the safe acceleration:
\begin{equation}
\label{eq: safe acceleration}
    a_{x,i}^{safe}(k+1) = 
\begin{cases} 
min \big( \frac{v^{diff}_i}{\Delta T} \hspace{0.2em}, \hspace{0.2em} A^-_x \big) & \text{if } v^{diff}_i < 0 \\
min \big(\frac{v^{diff}_i}{\Delta T} \hspace{0.2em}, \hspace{0.2em} A^+_x \big) & \text{otherwise}
\end{cases}
\end{equation}
where $A^+_x$ is the vehicle's maximum acceleration ability. Eq.~\ref{eq: safe acceleration} shows that a vehicle's acceleration and deceleration are constrained by its abilities, i.e., $A^-_x$ and $A^+_x$, respectively. 

During experiments for the current study, it was noticed that setting $A^-_x$ and $A^+_x$ according to maximum vehicle capabilities leads to highly unrealistic, sudden vehicle movements with very high jerk. Thus, their values are set to a reasonable value range for comfortable acceleration and jerk \cite{baeComfortableDrivingExperience2019}. Additionally, it was also noticed that after changing lateral positions, the new leader was sometimes not at a sufficient distance for vehicle $i$ to decelerate to the safe velocity without risking a collision, even when applying the desired deceleration $A^-_x$. To avoid this situation, if the braking distance is greater than the gap to the leader, i.e., $\frac{(v^{diff}_i)^2}{2A^-_x} > g_{leader}(k)-g_o$, the HDV uses a higher deceleration value $A^-_{critical}$ to avoid collision.

\subsubsection{Lateral Movement}

The lateral movements in the human model of \citet{mathewStripBasedApproachSimulation2015} are also based on strips rather than lanes. It is inspired by the lane change model of \citet{ehmanns2001simulationsmodell}, already implemented in SUMO \cite{krajzewiczTrafficSimulationSUMO2010}. \citet{mathewStripBasedApproachSimulation2015} applied this lane change approach to strips. The decision to change a strip is dependent on the benefit of changing the strip, measured in terms of speed gain. A vehicle is allowed to change only one strip in a time step; however, the model considers multiple strips in calculating the benefits since the driver's overtaking maneuver may require changing multiple strips. Thus, the decision to change the strip involves multiple steps. First, for a vehicle $i$, the benefit of changing the strip from a currently occupied strip $s_c$ to a destination strip $s_d$ is calculated using the following formula:

\begin{equation}
\label{eq: strip change benefit}
    b_{s_d, i} (k) = \frac{v^{safe}_{x, i, sd} (k) - v^{safe}_{x, i, sc} (k) }{v_{des,i}} \times e^{-\lambda \cdot n_s}
\end{equation}
where $v^{safe}_{x, i, sd}$ and $v^{safe}_{x, i, sc}$ are the safe velocities in strips $s_d$ and $s_c$, respectively. $v^{safe}_{x, i, sd}$ is calculated by imagining $i$ to be located at $s_d$ (instead of $s_c$), and calculating the leading vehicle and subsequently the safe velocity using Eq.~\ref{eq: safe velocity}. $n_s$ is the number of strip changes required to reach $s_d$ from $s_c$ and $\lambda$ is a model parameter to control the impact of far-away strips. Thus, the factor $e^{-\lambda \cdot n_s}$ is used to reduce the importance of benefits as the destination strip gets farther away from the currently occupied strip. In the denominator of Eq.~\ref{eq: strip change benefit},\citet{mathewStripBasedApproachSimulation2015} used the maximum speed possible in the current strip $s_c$. However, the current study replaced it with the desired speed of $i$. This modification is done for consistency with the PL controller for CAVs, which assumes that each CAV tries to achieve and maintain a specific desired speed. In simple words, Eq.~\ref{eq: strip change benefit} compares the potential speed in the destination strip to the speed possible in the current strip while considering the number of strip changes needed to reach the destination strip. If it is possible to gain speed in the destination strip, the benefits are positive; otherwise, they are negative.

With regards to the implementation of Eq.~\ref{eq: strip change benefit}, the model assumes that the decision to change the strip is not instantaneous; rather, the driver may keep track of the benefits for multiple time steps and only move when there is a significant accumulated benefit. Therefore, for each HDV, variables are maintained that represent the driver memory for observed benefits. The current study maintains separate variables for the accumulated benefits of strip changes to the left and right sides of the currently occupied strips. In each time step, Eq.~\ref{eq: strip change benefit} is calculated for all strips on the left and the right side, and accumulated to the respective memory variable; when the accumulated benefit crosses a certain threshold $L_{th}$, the driver changes the strip in the corresponding direction. The side that has the largest accumulated benefit is preferred for strip change. In order to avoid continuous changing of strips, the driver's memory variable is halved if the observed benefit on the corresponding side of the strip is less than or equal to zero. 

The value of $L_{th}$ models the aggressiveness of the human driver. A low value of $L_{th}$ would mean that the human driver makes lateral changes even for a slight gain in speed and vice versa for a high $L_{th}$. On the other hand, $\lambda$ determines how much changes in lateral position a human considers worth considering for the acquired benefits.

\subsection{Potential Lines Controller for LFT}

The PL controller introduced by \citet{rostami2023increasing} observes that the vehicles in countries with right-hand driving rules usually overtake and move faster on the left side of the road. PL controller utilizes this observation to laterally organize the LFT based on the desired longitudinal speeds of CAVs. The CAVs are assigned virtual PLs from left to right based on their desired speeds, as shown in Figure~\ref{fig: Lane Free Traffic Controller (b)}, encouraging the CAVs to follow the designated PL. This creates a lane-independent lateral structure and avoids unnecessary lateral movements.

For this study, any LFT controller could have been chosen to study the impacts of HDVs. However, the primary reason for choosing the PL controller is its simplicity and the fact that it does not require knowledge of the paths to be taken by other traffic participants, which is usually required for some optimization-based LFT approaches \cite{Karteek}. This requirement may be problematic for the inclusion of HDVs since the system does not exactly know the path that humans will take. However, the current study assumes that CAVs at least know the current speed of the HDVs, which can be estimated with high accuracy.

The PL controller of \cite{rostami2023increasing, potentialline} calculates the longitudinal acceleration based on two terms: (1) the goal to achieve the desired speed (the cruise control) and (2) the influential artificial forces of the other vehicles that prohibit collisions. These forces are also shown in Figure~\ref{fig: Lane Free Traffic Controller}. However, during experiments for the current study, it was noticed that artificial forces require significant parameter tuning to avoid collisions, which may still occur under high vehicle densities. Thus, the current study introduced the concept of safe velocity into the PL controller, inspired by the human driver model. Furthermore, \cite{rostami2023increasing, potentialline} used additional boundary forces using the proportional controller to force CAVs away from road boundaries. However, for consistency with the HDVs, the boundary forces are replaced by a hard constraint on the lateral acceleration that keeps the CAVs within road boundaries.

At first glance, the PL controller might appear similar to the strip-based HDV model described in the previous section. However, key differences distinguish the two. In the HDV model, strips are used solely to discretize the lateral axis, enabling vehicles to occupy positions more refined than traditional lanes, but overall, it still follows a conventional car-following modeling approach. Additionally, the resolution of the strips constrains lateral acceleration to discrete values, as vehicles are restricted to a single strip at any given time and can shift only one strip per time step.
In contrast, the PL controller does not rely on a car-following paradigm. Instead, it uses artificial potential fields to govern both lateral and longitudinal accelerations of CAVs. While these vehicles are encouraged to stay aligned with their assigned PLs, they are not strictly bound to them and can move freely along the lateral axis in response to the cumulative artificial forces. As a result, their lateral movements are significantly smoother than those of the HDV model.

With the above main distinguishing features, the following details the components used for the PL controller:

\subsubsection{PL based Lateral Control}

The PL controller uses a PL force to guide the vehicle to the assigned PL. As the first step, a lateral position $y_{pl,i}$, referred to as the PL, is assigned to the vehicle $i$ by linearly distributing the lateral axis according to the minimum and maximum desired speeds of all vehicles:

\begin{equation}
\label{eq: pl location}
    y_{pl,i} = Y_{r} + B_{pl} + (v_{des,i} - v_{min}) \frac{Y_{l} - Y_{r} - 2B_{pl}}{v_{max} - v_{min}} 
\end{equation}
where $Y_r$ and $Y_l$ are the lateral positions for the right and left boundaries of the road, respectively. The parameter $B_{pl}$ is used to leave some gap without PLs on either side of the road. This is required since the CAVs use the positions of their centers to align with the assigned PL. Therefore, $B_{pl}$ is determined based on the width of the broadest vehicle. $v_{max}$ and $v_{min}$ are the maximum and minimum desired speeds of all vehicles, respectively. 

After determining $y_{pl,i}$, the PL controller tries to steer and maintain the lateral position of the vehicle on the assigned $y_{pl,i}$ using a proportional controller:

\begin{equation}
    \label{eq: pl force}
    f_{pl, i} = K_{pl} \cdot (y_{pl,i} - y_i) - K_{pl,v} \cdot v_{y,i}
\end{equation}
where $K_{pl}$ and $K_{pl,v}$ are the controller gains for PL force.

\subsubsection{Cruise Controller based Longitudinal Control}

The fundamental aim of the cruise controller is to keep the longitudinal speed as close as possible to the desired speed. It achieves this using a proportional controller, given as:

\begin{equation}
    f^{cc}_{x,i} = K_{px} \big[v_{x,i}^{ts} - v_{x,i}(k) \big]
\end{equation}
where $K_{px}$ is the controller gain, allowing a gradual increase in speed. $v_{x,i}^{ts}$ is the target speed for the next time step, calculated using the vehicle's acceleration ability:

\begin{equation}
\label{eq: target speed}
    v_{x,i}^{ts} = min \{v_{x,i}(k) + A^+_x \cdot \Delta T \hspace{0.2em} , \hspace{0.2em} v_{des, i} \}
\end{equation}
where $A^+_x$ is the preferred acceleration. \citet{rostami2023increasing} did not use $A^+_x$ in their formulation. However, this study adds it to also cater for the situation when the vehicle comes to a complete halt. The current study uses $K_{px}$ value of 1.0, which makes the $f^{cc}_{x,i}$ to be directly determined by $A^+_x$ and $v_{des, i}$, in consistency with the human model (Eq.~\ref{eq: safe acceleration}).

\subsubsection{Potential Field based Collision Avoidance and Overtaking}

In an LFT, the collision avoidance and overtaking is generally performed via the artificial potential fields and the resulting inter-vehicular forces \cite{yanumulaOptimalPathPlanning2021}. Each surrounding vehicle is considered a moving obstacle, which the subject vehicle sees as an ellipsoid hemisphere. Figure~\ref{fig: Lane Free Traffic Controller (a)} shows an example of the potential field of vehicle V1 (as seen by other CAVs) and corresponding artificial forces as experienced by vehicles V2 and V3.

\citet{rostami2023increasing} used a modified version of the original LFT forces formulation of \cite{yanumulaOptimalPathPlanning2021}. However, during experiments for the current study, no significant difference was observed in the performance of the two formulations. With correct parameter tuning, both formulations performed quite similarly. Therefore, the current study uses the original formulation of \cite{yanumulaOptimalPathPlanning2021} for simplicity. For a vehicle $j$ in the surrounding of subject vehicle $i$, the artificial force is calculated as:

\begin{equation}
\label{eq: lft force}
    F^{ij} = \frac{1}{ \left[ \left( \frac{x_i - \delta_{ij}}{0.5 s_{d_1}} \right)^{p_1} + \left( \frac{y_i - y_j}{0.5 s_{d_2}} \right)^{p_2} \right]^{p_3} + 1  }
\end{equation}
where $p_1$, $p_2$, and $p_3$ are the function parameters, and $s_{d_1}$ and $s_{d_2}$ determine the longitudinal and lateral axis of the ellipsoid, respectively. $\delta_{ij}$ is a function for adjusting the longitudinal position of the ellipsoid's center for vehicle $j$, considering the safety gap and speeds (refer to \cite{yanumulaOptimalPathPlanning2021} for details). The parameters $p_1$, $p_2$, and $p_3$ were set as 2, 2, and 6, respectively \cite{yanumulaOptimalPathPlanning2021}.

The calculated $F^{ij}$ is projected to the longitudinal and lateral axes, creating two components: $F^{ij}_x$ and $F^{ij}_y$. These two components are accumulated for all the vehicles up to distance $\Delta D_{front}$ in the front and $\Delta D_{back}$ in the back, forming the nudging and repulsive forces, respectively.

\subsubsection{Overall Accelerations}

After calculating the cruise controller and artificial forces, the PL controller calculates the longitudinal and lateral accelerations as:
\begin{align}
    a^{pl}_{x,i}(k+1) &= f^{cc}_{x,i} + w_{n} \sum_{j \in V_{front}} F^{ij}_x + w_{r} \sum_{j \in V_{back}} F^{ij}_x \\
    a_{y,i}(k+1) &= w_{n} \sum_{j \in V_{front}} F^{ij}_y + w_{r} \sum_{j \in V_{back}} F^{ij}_y + f_{pl, i} \label{eq: CAV Lateral ACC}
\end{align}
where $w_{n}$ and $w_{r}$ are the weights for nudging and repulsive forces, respectively. $V_{front}$ and $V_{back}$ are the set of all vehicles (including HDVs) in front and back of the vehicle $i$ up to distance $\Delta D_{front}$ and $\Delta D_{back}$, respectively. Even though it is possible to use different values for $w_{n}$ and $w_{r}$, leading to significantly different LFT behavior \cite{rostami2023increasing, yanumulaOptimalPathPlanning2021}, for simplicity, the current work uses a value 1.5 for both parameters.

\subsubsection{Boundary and other constraints for Lateral Acceleration}

The lateral acceleration obtained in Eq.~\ref{eq: CAV Lateral ACC} must be restricted to keep the vehicles within the road boundary. A boundary feedback controller is used for this purpose to smoothly limit the lateral acceleration such that the vehicles do not cross the road boundaries \cite{malekzadeh2022empirical, potentialline}. It is calculated as:
\begin{equation}
    a_{y,i}^{lim} = K_{b1} \cdot (\hat{y} - y_i) - K_{b2}\cdot v_{y,i}
\end{equation}
where $\hat{y}$ is the road boundary limits that $y_i$ must not cross. Let $W_i$ represent the width of the vehicle $i$. Since $y_i$ is the lateral position of the center of the vehicle, setting the value of $\hat{y}$ to $Y_l - 0.5W_i$ and $Y_r + 0.5W_i$ determines the upper and lower limits for the lateral acceleration such that the vehicle does not cross the road boundary.

In addition to boundary control, the lateral acceleration is also bounded by minimum and maximum lateral acceleration ($A_y^{min}$ and $A_y^{max}$) and jerk ($J_y^{min}$ and $J_y^{max}$) limits for comfortable accelerations.

\subsubsection{Extension of Collision Avoidance via Safe Acceleration}
\label{section: Safe Acceleration in LFT}

So far, the PL components described above have been taken from the literature without significant modification. In contrast, the following describes an essential modification introduced by the current study.

Even though the PL controller already considers artificial forces to avoid collisions, it is observed that these forces may cancel each other out under certain conditions. This causes the vehicle to not decelerate on time, leading to collisions, especially at high vehicle densities. This can be avoided by adjusting the individual weights given to different forces; however, tuning these parameters is a time-consuming process and may not necessarily produce a general parameter set suitable for all situations.

In view of the above, the current study introduces the constraint of safe acceleration into LFT controller. The first step in this process is selecting a leader vehicle. Similar to the human model, the vehicle with the shortest longitudinal distance and an overlap in the lateral axis is chosen as the leader for vehicle $i$. However, unlike the human model, the overlap is directly calculated using vehicle widths and lateral positions without discretizing the lateral axis into strips. After determining the leader, the safe velocity $v^{safe}_{x, i}$ and the safe acceleration $a_{x,i}^{safe}$ are calculated using Eq.~\ref{eq: safe velocity} and Eq.~\ref{eq: safe acceleration}, respectively. $a_{x,i}^{safe}$ is then used to bound the longitudinal acceleration calculated by the PL controller; thus, the longitudinal acceleration is given as:
\begin{equation}
    a_{x,i} (k+1) = min\{a^{pl}_{x,i}(k+1) \hspace{0.2em}, \hspace{0.2em} a_{x,i}^{safe}(k+1)\}
    \label{eq: PL Safe condition}
\end{equation}

Finally, the $a_{x,i} (k+1)$ is also bounded by lower and upper acceleration ($A_x^{min}$ and $A_x^{max}$) and jerk ($J_x^{min}$ and $J_x^{max}$) limits, similar to lateral acceleration. It is important to note that even though the cruise controller in Eq.~\ref{eq: target speed} uses the acceleration $A^+_x$ to achieve the desired speed, the actual acceleration of CAVs can be higher than $A^+_x$ due to combined nudging and repulsive forces of surrounding vehicles. This is in contrast to the HDV model, where no such forces exist to cause higher longitudinal acceleration than $A^+_x$. Thus, it is essential to bound the longitudinal acceleration of CAVs, which allows fine-tuning the rest of the LFT parameters for comfortable LFT movements.

\subsection{Adaptive Potential Lines}

\begin{figure}[tb]
     \centering
     \begin{subfigure}[b]{.49\textwidth}
         \centering
         \includegraphics[width=1\textwidth]{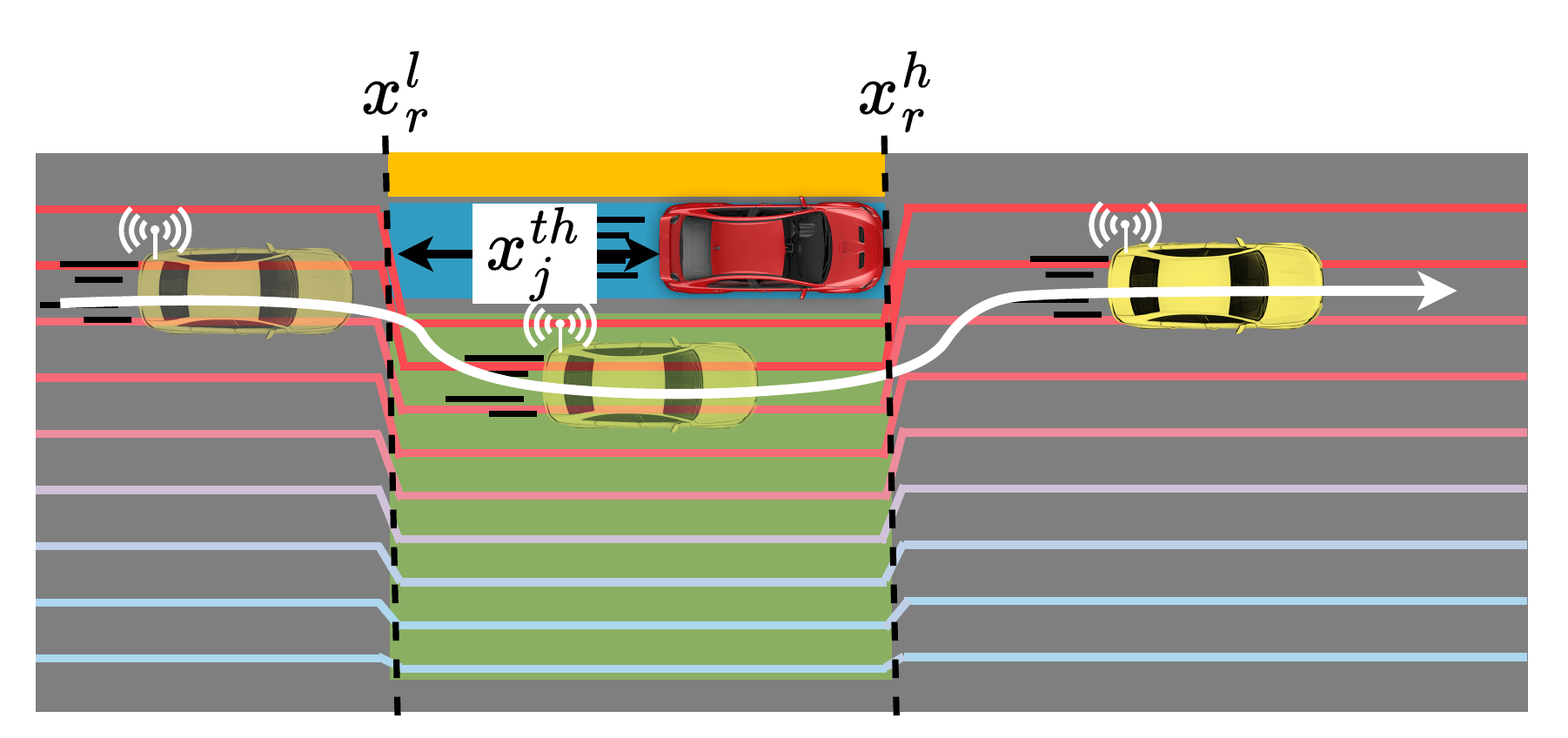}
         \caption{Single HDV case.}
         \label{fig: adaptive pl (a)}
     \end{subfigure}
     \hfill
     \begin{subfigure}[b]{.49\textwidth}
         \centering
         \includegraphics[width=1\textwidth]{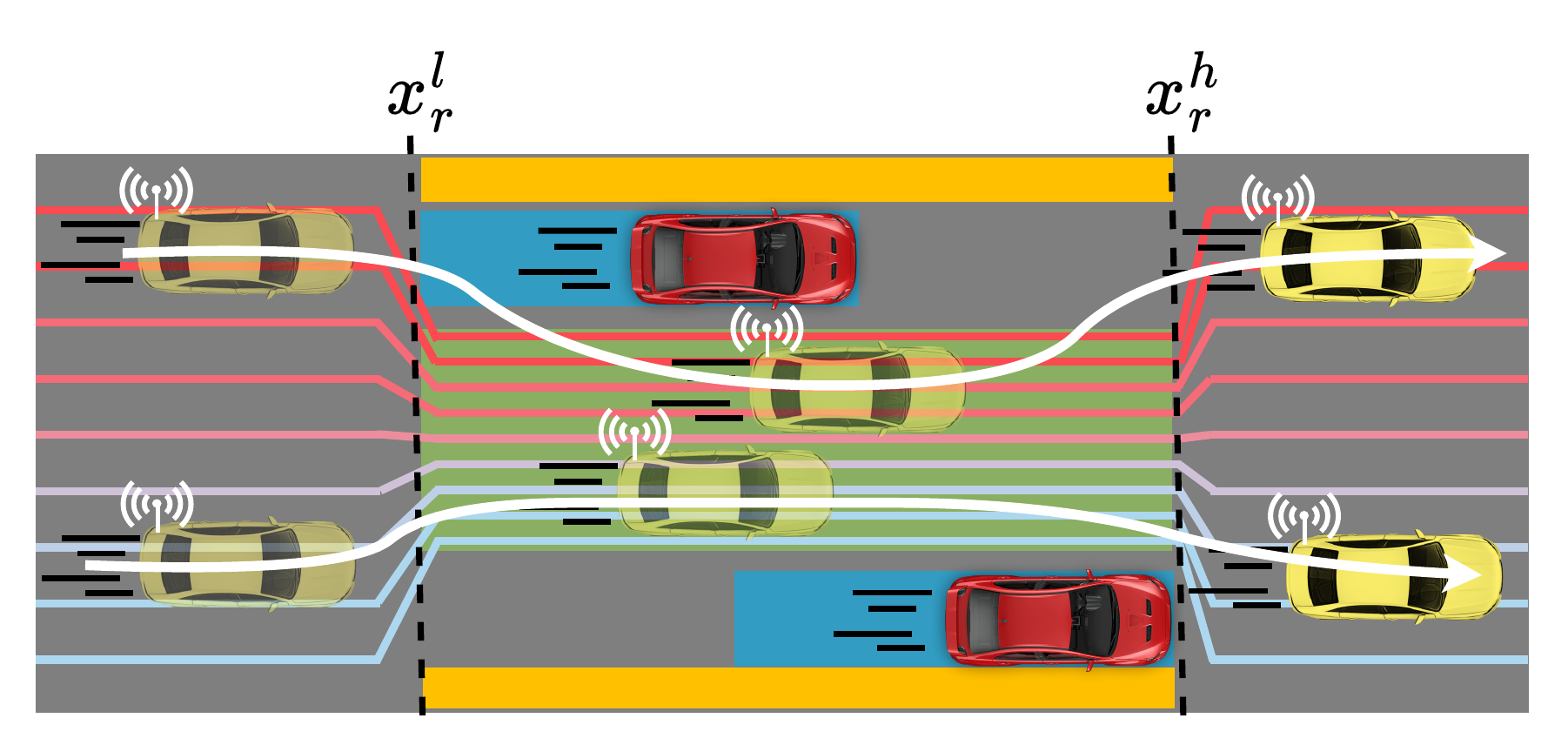}
         \caption{Laterally continuous APL corridor.}
         \label{fig: adaptive pl (b)}
     \end{subfigure}
     \hfill
    \begin{subfigure}[b]{.49\textwidth}
         \centering
         \includegraphics[width=1\textwidth]{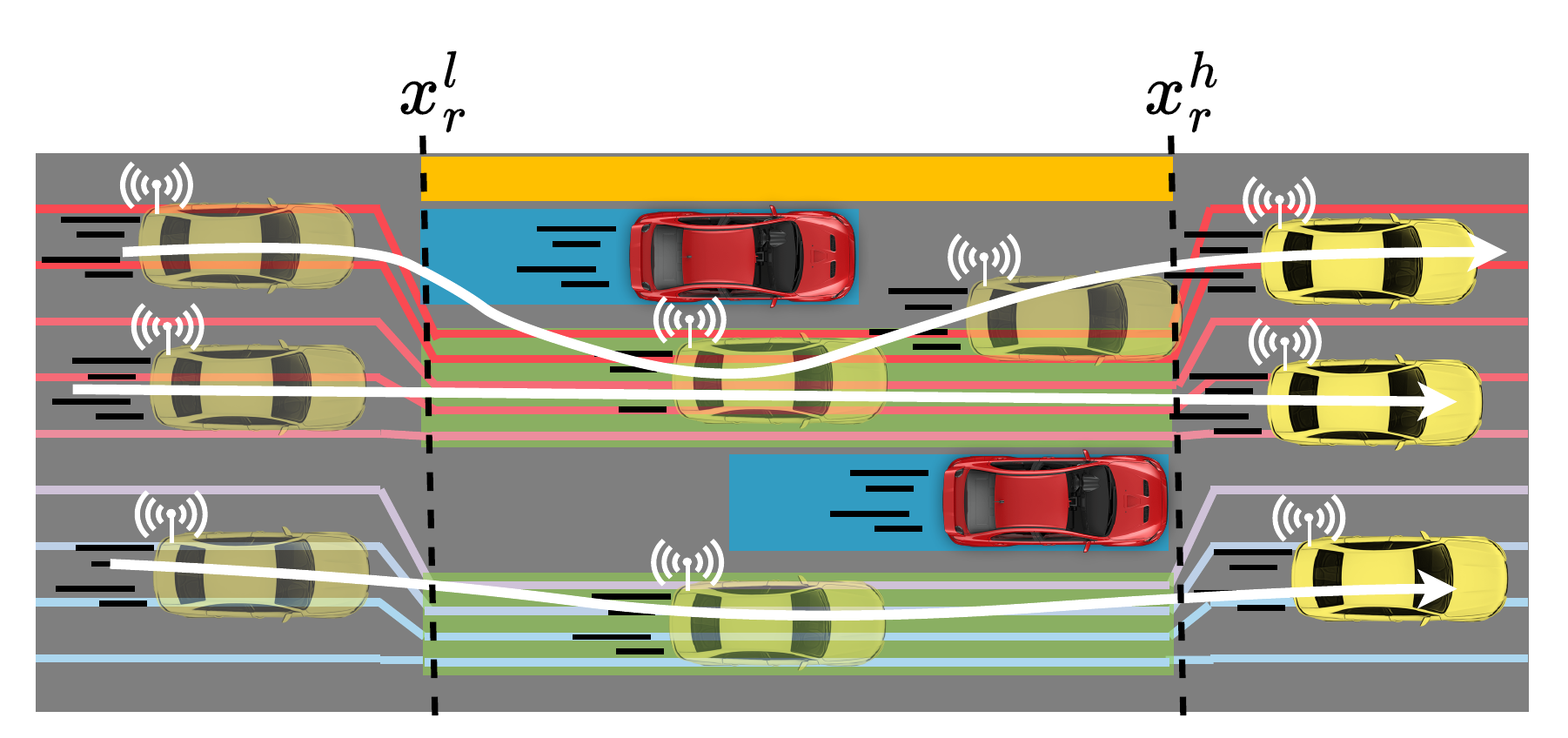}
         \caption{Laterally discontinuous APL corridor.}
         \label{fig: adaptive pl (c)}
     \end{subfigure}
        \caption{Adaptive potential lines (APL) controller. It adapts the PL areas near HDVs, represented by the green color and referred to as APL corridor in the study. The yellow color shows the areas excluded from PLs due to insufficient space to fit a vehicle. The blue color shows the areas marked as occupied by HDVs.}
        \label{fig: adaptive pl}
\end{figure}

The presence of HDVs can degrade the performance of a LFT controller. This degradation primarily occurs because CAVs in LFT cannot exert artificial forces on HDVs, which restricts the free movement of CAVs. This will eventually affect the performance of each LFT controller differently. However, for the PL controller, a significant restriction in CAV movement can occur when HDVs drive on PLs assigned to CAVs. In such cases, if a CAV is following an HDV and is already on the assigned PL, there will be no PL force ($f_{pl,i}$) to steer the CAV away from the PL and overtake the slow-moving HDV ahead, causing a cumulative effect on upstream vehicles.

According to the functioning of the LFT controller, each controller must devise a strategy to cater to the HDVs. Since this study uses a PL controller to study the effects of HDVs on LFT, it also modifies the PL controller to lower the impacts of HDVs on LFT. Accordingly, this study introduces the concept of APL controller, in which the PLs are modified in the vicinity of HDVs. This section outlines the APL strategy for LFT, beginning with a general description of how the vicinity of modified PLs around HDVs is calculated. It then discusses the various methods tested in the study for activating APL around HDVs. These two steps are calculated and applied in each simulation time step.

\subsubsection{Adaptive Potential Lines (APL) Corridors}

The APL controller primarily compresses the PLs into the lateral spaces between HDVs, as shown in Figure~\ref{fig: adaptive pl}. This adjustment enables CAVs to overtake slow-moving HDVs and enhances the overall flow of CAVs. Since CAVs already move in a coordinated manner due to the PL controller, they can also overtake HDVs in a coordinated way by simply modifying the assigned PL position ($y_{pl,i}$) around HDVs. Additionally, these areas can create long road regions where CAVs are longitudinally dominated, allowing them higher chances to exert artificial forces on the downstream CAVs and accelerate without being blocked by HDVs. This functions as a sort of corridor for CAVs with comparatively higher freedom and, thus, is referred to as the APL corridor in the study.This study assumes that APL calculations are performed centrally by dedicated controllers positioned along the roadside. These controllers estimate the locations and speeds of HDVs and transmit the relevant APL information to CAVs. Alternatively, a decentralized approach could be employed, where individual CAVs contribute to the APL calculations. In this scenario, CAVs would share their location and speed estimates with one another via V2V communication, allowing a distributed algorithm to fuse the data for improved accuracy in HDV estimation. The fused data would then be used to perform the APL calculations and disseminate the resulting information to nearby CAVs.

%This modification allows CAVs to overtake slow-moving HDVs and improve the overall flow of CAVs, as shown in Figure~\ref{fig: adaptive pl (a)}. 

To determine the areas where APL corridors can be formed, the method first divides the road into a set of regions $R$ in the longitudinal axis, with $x^l_r$ and $x^h_r$ denoting the longitudinal positions of the start and end of the region $r \in R$, respectively. A region starts at a distance $x^{th}_j$ behind the HDV $j$ and extends to its front in the longitudinal axis. $x^{th}_j$ provides additional safety distance for the CAVs to orient themselves on the modified PLs without excessive deceleration near the HDVs. This region does not extend beyond the front of the HDV to avoid influencing CAVs in front, which is problematic since any HDV approaching from behind would force the CAV to move away from its front, giving unnecessary priority to HDVs. If multiple regions overlap in the longitudinal direction, they are merged to form a single region --- an extended APL corridor. Figures~\ref{fig: adaptive pl (b)} and \ref{fig: adaptive pl (c)} show examples of such longitudinal overlap and extended APL corridor.

After determining the longitudinal regions, the method divides the lateral axis of each region. For this purpose, combinations of lateral positions are computed for the stretch of the road from $x^l_r$ to $x^h_r$. These combinations of lateral positions remain fixed for the entire region. They are calculated by leaving a lateral gap of $B_{apl}$ on either side of HDVs. Similar to the PL formulation in Eq.~\ref{eq: pl location}, this gap is necessary since the CAVs use center positions to align themselves to the assigned PLs. $B_{apl}$ is determined by half of the width of the broadest vehicle. Without $B_{apl}$, the overtaking maneuver of the CAVs may be restricted due to lateral overlap with HDVs and the constraint of safe acceleration. Furthermore, if the lateral gap between two HDVs is less than 2$B_{apl}$ (meaning a CAV may not fit into the gap), the corresponding lateral gap is not considered for APL corridor. Thus, within a single longitudinal stretch ($X_l$, $X_h$), the APL corridor can be laterally continuous (Figure~\ref{fig: adaptive pl (b)}) or discontinuous (Figure~\ref{fig: adaptive pl (c)}), depending on the situation. If there is not enough lateral gap in the region to fit even a single CAV, then the region is not considered for an APL corridor. 

\subsubsection{Activation Condition for APL Corridors}

The previous section focused on the potential areas for APL corridors. The two critical factors for applying APL corridors are (1) determining $x^{th}_j$ and (2) determining the conditions needed for an HDV to include its surroundings into APL corridors. The paper evaluates four approaches that mainly differ in these two aspects. Figure~\ref{fig: apl methods} shows the parameters used by these methods. The APL methods used are described below:

\begin{figure}[!tb]
     \centering
         \includegraphics[width=0.6\textwidth]{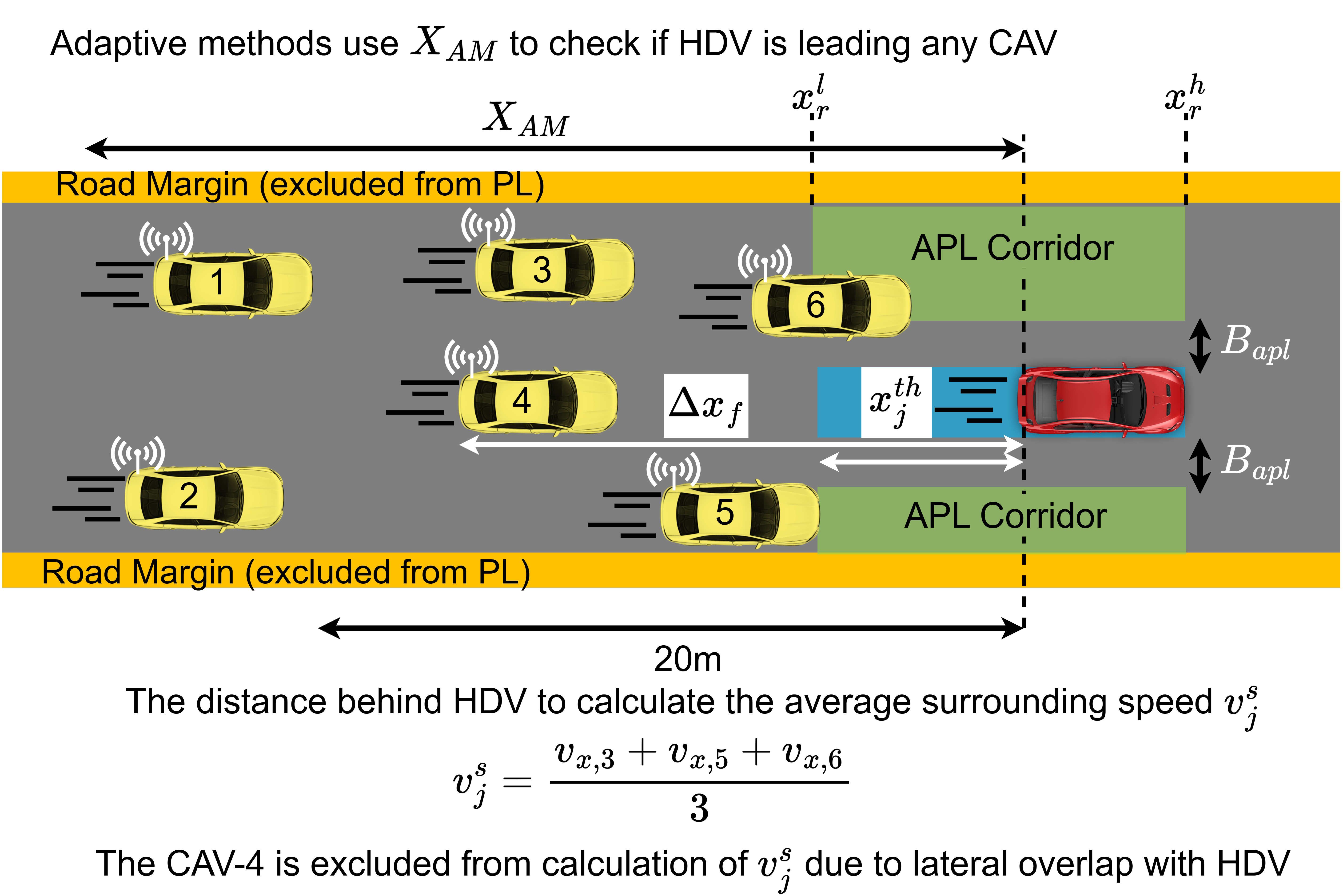}
        \caption{Schematic to show parameters used for APL methods.}
        \label{fig: apl methods}
\end{figure}

\begin{table}[]
\centering
\begin{tabular}{@{}lp{4cm}l@{}}
\toprule
\textbf{Method} & \textbf{The value of $x_j^{th}$} & \textbf{Condition for Corridor   Formation}                           \\ \midrule
CM              & Fixed as $X_{CM}$                & Unconditional                                                         \\
NSCM            & Fixed as $X_{CM}$                & HDV's speed is less than   surrounding speed, i.e. $v_{x,j} < v_j^s$. \\
FAM &
  Dependent on the distance of the CAV following HDV, i.e., $\Delta x_f$ &
  \begin{tabular}[c]{@{}l@{}}(1)   $v_{x,j} < v_j^s$ \\ (2) The HDV leads a CAV within $X_{AM}$ distance.\end{tabular} \\
SVAM &
  Dependent on the distance   of the CAV following HDV, i.e. $\Delta x_f$ &
  \begin{tabular}[c]{@{}l@{}}(1) $v_{x,j} < v_j^s$ \\ (2) The   HDV leads a CAV within $X_{AM}$ distance. \\ (3) The follower CAV's velocity   restricted by the \\ safe velocity, i.e., $v_{x,i} < 1.05 v_{x,i}^{safe}$\end{tabular} \\ \bottomrule
\end{tabular}
\caption{Summary of the APL methods.}
\label{table: APL summary}
\end{table}

\begin{enumerate}
    \item \textbf{Constant Margin (CM)}: This is the simplest method. It keeps $x^{th}_j$ constant (denoted as $X_{CM}$) and continuously uses all HDVs' vicinities to define APL corridors.

    \item \textbf{Neighbouring Speed based Constant Margin (NSCM)}: Similar to CM method, this method uses the same $X_{CM}$ for all HDVs; however, the vicinity (i.e. longitudinal distance $X_{CM}$ behind an HDV) of an HDV is only included into APL corridor if its speed is less than the average speed of the surrounding vehicles. For an HDV $j$, the average surrounding speed, denoted by $v^s_j$, is computed by taking the mean longitudinal speed of all vehicles (both CAVs and HDVs) whose positions do not laterally overlap with the boundaries of $j$ and are within a longitudinal distance of 20~m behind $j$. Figure~\ref{fig: apl methods} provides an example of calculating $v^s_j$. The idea here is that the APL should only be applied in the surroundings of an HDV if it is blocking the vehicles behind it, which can be detected if the vehicles with different lateral positions than $j$ are moving faster than $j$, i.e., $v_{x,j} < v_j^s$.

    \item \textbf{Follower-based Adaptive Margin (FAM)}: The PL controller in this study uses the leader-follower relation along with safe acceleration as a hard constraint to avoid collisions (Eq.~\ref{eq: PL Safe condition}). The HDV can potentially hinder any CAV that has an HDV leader. FAM uses this characteristic to apply the APL. It first checks if the HDV $j$ is the leader of any CAV within a longitudinal distance of $X_{AM}$ behind the HDV. Then, it checks if the speed of $j$ is lower than the average speed of the surrounding vehicles (the same as in the NSCM strategy). If both conditions are fulfilled, $x^{th}_j$ is set as the distance from the back of HDV to the back of the follower CAV, shown as the distance $\Delta x_f$ in Figure~\ref{fig: apl methods}. The purpose here is to only apply APL behind an HDV up to a point that allows the overtaking of the follower CAV. Thus, $x^{th}_j$ in this strategy is not constant; rather, it depends on the distance of the follower CAV, as reflected in its name. 

    \item \textbf{Safe Velocity based Adaptive Margin (SVAM)}: This approach extends the FAM method. In addition to checking if HDV is the leader of a CAV, it also checks if the speed of the follower CAV is constrained by the safe velocity, i.e., $v_{x,i} \leq (1+\epsilon) v^{safe}_{x,i}$, where $i$ is the follower CAV and $\epsilon$ is a small number, set to 0.05 in the study. The main idea here is to activate APL only if the follower CAV's speed is limited due to HDV. The rest of the conditions of SVAM are the same as FAM.
    
\end{enumerate}

Table~\ref{table: APL summary} summarizes the parameters and conditions used by each APL method.

\section{Experimental Setup}

To study the effects of HDVs, a ring road of 1~$km$ is simulated in a microscopic simulation. A custom extension of SUMO \cite{behrisch2011sumo} for LFT called TrafficFluid-Sim \cite{troullinosLaneFreeMicroscopic2021} is used for this purpose. The HDVs and PL controller are implemented via a C++ interface, which allows getting the necessary information on individual vehicles and setting up accelerations for the next time step. For consistency with other LFT works, the road width is set to 10.2~m. The road emulates a continuous beltway, with the vehicles leaving from one end and entering back into the road from the other with the same lateral positions and speeds. 

Regarding the simulation scenarios, five types of vehicles are simulated with (length, width) in meters given as (3.2, 1.6), (3.4, 1.7), (3.9, 1.7), (4.55, 1.82), and (5.2, 1.88). The experiments simulate vehicle densities ranging from 50 $veh/km$ to 400 $veh/km$ with a step size of 50 $veh/km$, with an equal proportion of vehicles out of the five above-mentioned categories. Different penetration rates of HDVs are simulated. If the penetration rate leads to a fractional outcome for the number of HDVs, it is rounded to the closest integer. Using uniform distribution, the vehicles are initialized at random positions with zero speed without vehicle overlap. The desired speeds are also assigned using uniform distribution ranging from 25~$m/s$ to 35~$m/s$, which remain constant throughout the simulation. Each scenario is run for one hour of simulation time with a step size $\Delta T$ of 0.25~s. Each scenario is also run with five random seeds to ensure the results are statistically reliable.

The following values are used for the model parameters. For the human model, $\Delta s$, $\lambda$ and $L_{th}$ are set to 0.05~$m$, 0.1, and 10, respectively. The values of $\Delta s$ and $\lambda$ are based on the values used in \cite{mathewStripBasedApproachSimulation2015}, while $L_{th}$ is not explicitly mentioned by the authors, and thus, the value is chosen by experiments to allow sufficient strip changing maneuvers. For calculating $v^{safe}_{x, i}$, the deceleration $A^-_x$ is set as -1.5~$m/s^2$ to ensure smooth deceleration and minimize jerk. To account for situations requiring more aggressive braking of HDVs due to changes in lateral positions, the higher deceleration threshold $A^-_{critical}$ is set as -2.6~$m/s^2$. Note that $A^-_{critical}$ is only used by HDV as the LFT forces already cater for it for CAVs. $\tau$ is fixed as 0.5~$s$ for all CAVs. However, to bring variety to the individual driving style of each human driver, for HDVs, $\tau$ is drawn from a normal distribution with a mean and standard deviation of 1.5 and 0.5, respectively. The minimum safety gap $g_o$ is set to 2~m. The desired acceleration $A^+_x$ of all vehicles is also set to 1.5~$m/s^2$. For CAVs, the acceleration limits $(A_x^{min}, A_x^{max})$ are set to (-4.5~$m/s^2$, 2.6~$m/s^2$) and $(A_y^{min}, A_y^{max})$ are set to (-1.5~$m/s^2$, 1.5~$m/s^2$). The jerk limits $(J_x^{min}, J_x^{max})$ and $(J_y^{min}, J_y^{max})$ are set to (-2.0~$m/s^3$, 2.0~$m/s^3$). The acceleration and jerk values are mainly derived from \cite{baeComfortableDrivingExperience2019} for comfortable and safe driving experiences. $\Delta D_{front}$ and $\Delta D_{back}$ are set to 100~$m$. The weights for nudging and repulsive forces, i.e., $w_n$ and $w_r$, are set as 1.0 and 0.5, while the controller gains for the PL controller $K_{pl}$ and $K_{pl,v}$ are set as 0.02 and 0.65, respectively. For boundary control, the values of $K_{b1}$ and $K_{b2}$ are set as 4 and 3.75, respectively. $X_{CM}$ and $X_{AM}$ are set to 40~$m$ by default unless explicitly stated otherwise.

\section{Results and Discussion}

The results are divided into two main sections. The first section discusses the effects of increasing HDVs in LFT. The second section then analyses how this performance drop can be improved by using the APL controller.

\subsection{Impact of HDVs on the LFT performance}

\begin{figure}[!tb]
  \centering
    \includegraphics[width=0.82\linewidth]{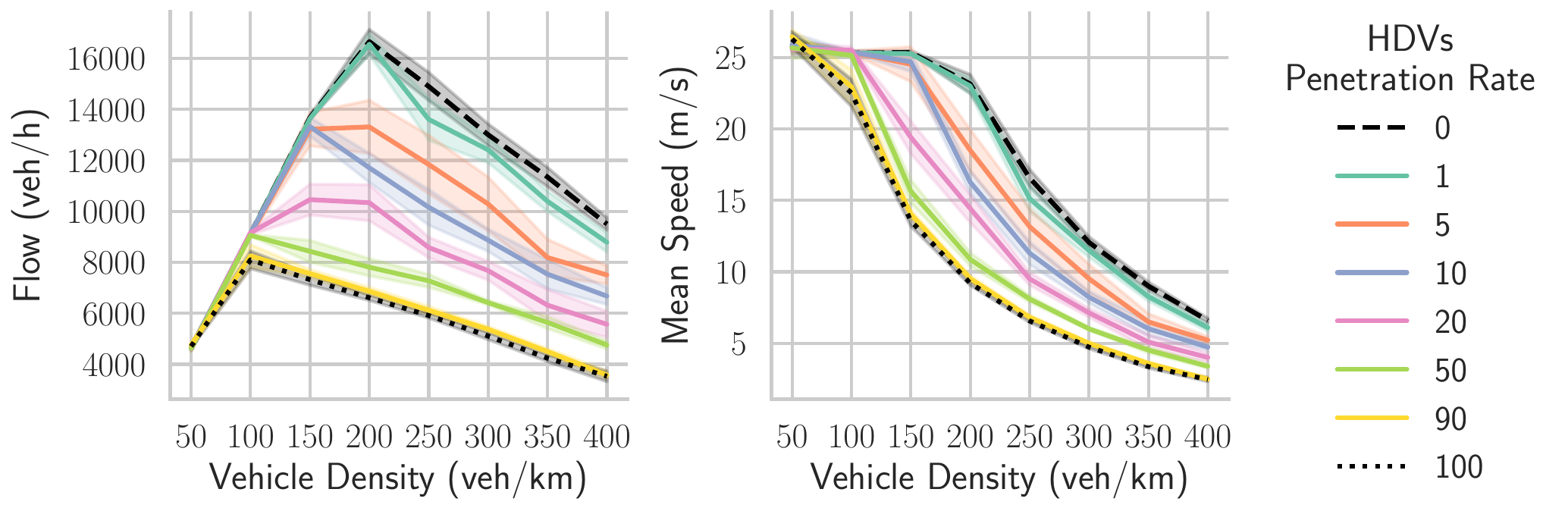}
  \caption{Traffic flow rate and mean speed of PL controller with different HDV penetration rates. The 0\% and 100\% penetration rates mark all-CAVs and all-HDVs scenarios, respectively. The shaded area shows the standard deviation.}
  \label{fig: flow and speed}
\end{figure}

The study first analyses the impact of different penetration rates of HDVs on the overall LFT performance, especially in terms of traffic flow and average speed, as shown in Figure~\ref{fig: flow and speed}. The first thing to observe is the significant difference in the road capacities (the highest traffic flow in Figure~\ref{fig: flow and speed}) with all-HDVs and all-CAVs scenarios. With 100\% HDVs, the road capacity is limited to only 8,100~$veh/h$ (speed: 22.5~$m/s$) at a density of only 100~$veh/km$, which is roughly equivalent to the flow of lane-based traffic with four lanes (approximately 2000~$veh/h/lane$ \cite{kesting2013traffic}). Notably, the road width used in the simulation (10.2 meters) is typically divided into three lanes in lane-based traffic management. However, since the lanes are usually wider than the vehicles, removing the lanes allowed four vehicles to fit within the same road width, resulting in a flow roughly equivalent to four lanes. Nevertheless, this flow value also depends on the tolerance of individual drivers for changing lateral positions. For instance, in the experiments conducted for this study, the maximum road capacity with all-HDVs decreased from 8,100~$veh/h$ to 7,400~$veh/h$ when the threshold for benefits of changing strip ($L_{th}$) was reduced from 10 to 0.1.

In contrast, the scenario with all-CAVs achieves a significantly higher road capacity of 16,700~$veh/h$\footnote{The 16,700~$veh/h$ road capacity is observed using the baseline parameter values mentioned in the paper. A higher flow is still possible with further tuning of the parameters. For example, by setting $w_n$ and $w_r$ to 1.5 and 1.0, respectively, a road capacity of 18,000~$veh/h$ was observed. However, this led to significantly higher CAV movements under mixed scenarios with HDVs. The road capacity increases even further by adjusting the minimum safety gap $g_o$ and reaction time $\tau$. For example, it increases to almost 20,000~$veh/h$ by setting $g_o$ to 0, of course, at the cost of significantly higher safety risks.} (speed: 23.1~$m/s$) at a density of 200~$veh/km$ --- an increase of almost 106\% in road capacity over all-HDVs scenarios. The key difference between HDVs and the LFT is the existence of artificial forces, indicating that the coordinated movement of CAVs through artificial forces and virtual PLs plays a crucial role in achieving higher flow. Additionally, despite using the same desired speeds for individual vehicles in both scenarios, the average speed observed in the all-CAVs scenario is significantly higher across all vehicle densities.

It is also worthwhile to compare the obtained maximum capacity with conventional lane-based traffic. For a 10.2~m three-lane road, conventional lane-based traffic reaches about 6,000~$veh/h$, whereas lane-based CAVs with Cooperative Adaptive Cruise Control (CACC) could theoretically achieve up to 12,000~$veh/h$ (assuming 3500–4000~$veh/h/lane$) under ideal conditions \cite{papamichail2019motorway,shladover2012impacts,vander2002effects}. This remains well below the 16,700~$veh/h$ observed for all-CAV LFT. In practice, lane-based CACC capacity would be even lower once lane-changing disruptions are considered, whereas LFT avoids these losses through continuous lateral positioning.

\begin{figure}[!tb]
  \centering
    \includegraphics[width=1\linewidth]{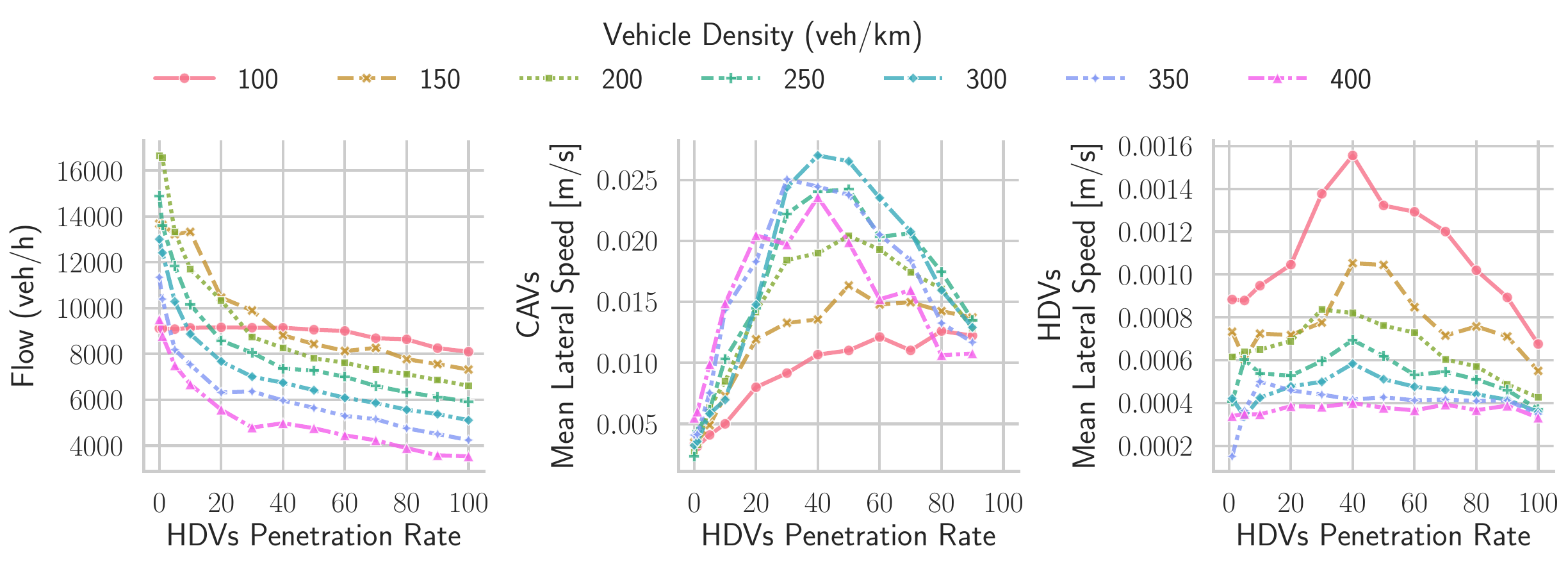}
  \caption{Traffic flow rate and mean lateral speeds for increasing penetration of HDV into LFT.}
  \label{fig: relative flow and speed}
\end{figure}

\begin{figure}[!tb]
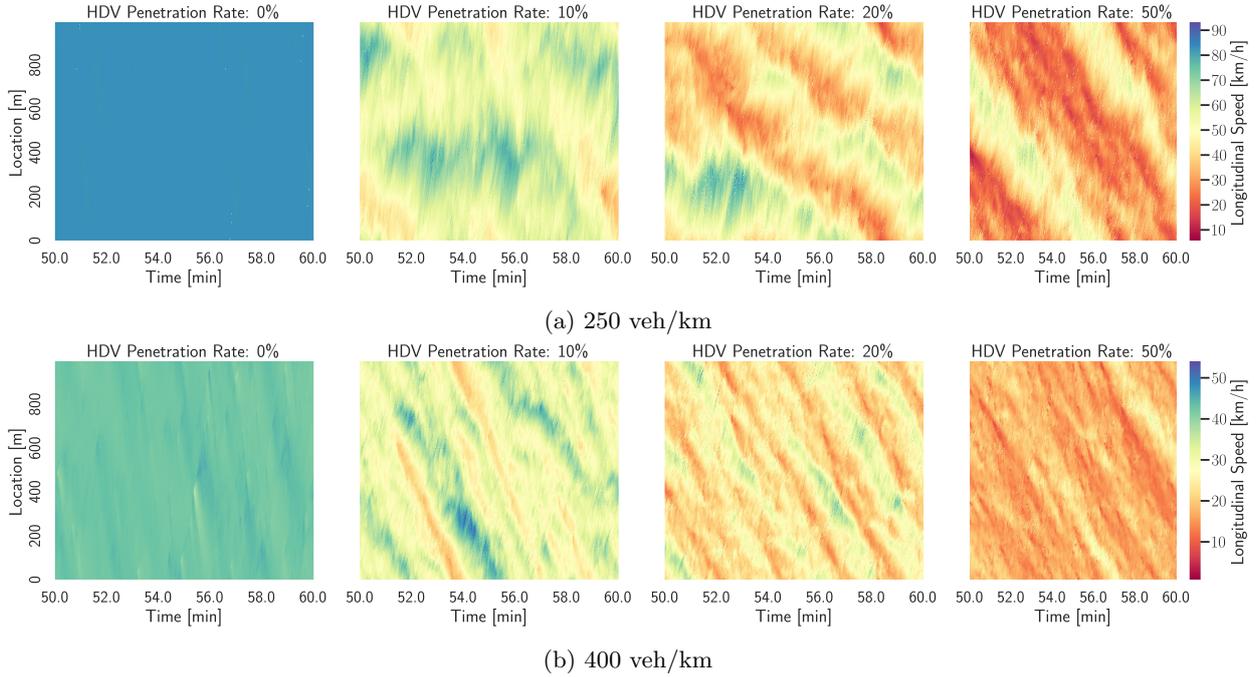

     \centering
     \begin{subfigure}[b]{1\textwidth}
         \centering
         \includegraphics[width=1\textwidth]{figures/trc_second_revision_new/waves_heatmap_pl_200_seed_2_nudge_1.jpg}
         \caption{200~veh/km}
         \label{fig: space time (a)}
     \end{subfigure}
     \hfill
     \begin{subfigure}[b]{1\textwidth}
         \centering
         \includegraphics[width=1\textwidth]{figures/trc_second_revision_new/waves_heatmap_pl_400_seed_2_nudge_1.jpg}
         \caption{400~veh/km}
         \label{fig: space time (b)}
     \end{subfigure}
        \caption{Spatio-temporal speed plot for PL controller. The formation of traffic waves is visible with the increasing penetration of HDVs.}
        \label{fig: space time}
\end{figure}

Figure~\ref{fig: flow and speed} also shows that an extremely small HDV penetration of 1\% (i.e., 1-4 HDV depending on vehicle density) has a comparatively limited effect on the LFT flow. This HDV penetration is relevant to scenarios where some exceptional vehicles, such as emergency vehicles or towed cars, have to be manually driven. However, with even slight increase in HDV penetration, the LFT flow is severely affected: with just a 5\% penetration rate, road capacity drops to  13,300~$veh/h$ (a drop of nearly 20\%), and with a 10\% penetration rate and the same vehicle density of 200~$veh/h$, the flow is decreased to 11,700~$veh/h$ (a drop of nearly 30\%). Interestingly, at 10\% HDVs the maximum capacity remains nearly the same as with 5\% HDVs, but the critical density decreases to 150~$veh/km$. This shift occurs because higher HDV penetration reduces average traffic speed, limiting CAVs’ ability to overtake or influence HDV behavior. Nevertheless, the PL controller still maintains similar capacity at 10\% HDVs as at 5\%, although at the lower critical density. This demonstrates that HDV penetration not only degrades LFT performance but can also shift the fundamental diagram, depending on CAV–HDV interaction. This phenomenon is further analyzed in Figure~\ref{fig: relative flow and speed}, illustrating traffic flow against increasing HDVs penetration rates. It is found that the above drastic traffic flow drop with 5\% and 10\% HDVs penetration is consistent for all vehicle densities above $150~veh/km$. In fact, the performance drop is significantly high for the initial 0-20\%, where at 20\% penetration rate, the flow already drops by 40\%. Beyond this point, the curves are almost flat with limited performance gain over all-HDVs scenarios: from 40-100\% the performance drops further by only 10\%, reaching a drop of nearly 60\% for all-HDVs scenario. %The performance decline is more pronounced at higher vehicle densities.

The above phenomena can be better understood by examining vehicle movements along the lateral axis. In general, when slower vehicles do not obstruct faster ones, vehicles with higher desired speeds maintain their trajectory with minimal lateral deviation, resulting in smoother traffic flow. Conversely, when slower vehicles impede faster vehicles, the latter must frequently adjust their lateral position to overtake, leading to increased lateral movement. We quantify this behavior using the mean of the absolute values of lateral speeds, referred to as the mean lateral speed in this study. As shown in Figure~\ref{fig: relative flow and speed}, in the all-CAVs scenario, vehicles are able to better coordinate their movements and remain on their respective PLs, resulting in reduced lateral movement and a correspondingly lower mean lateral speed. For all-CAVs scenarios, it is also observed that the vehicles generally show higher lateral movements at lower vehicle densities compared to higher vehicle densities. This can be explained by the higher available space at lower densities, causing the overtaking vehicles to fully utilize their artificial forces to nudge the vehicles in the front. As the vehicle densities increase, the nudging possibility decreases due to the artificial forces from the surrounding vehicles countering each other. Nonetheless, the vehicles can still coordinate their movements at higher densities, and flow remains significantly higher than all-HDV scenarios.

As the proportion of HDVs increases, lateral movements rise significantly. CAVs find it increasingly difficult to nudge other vehicles, as nudging does not affect HDVs, resulting in continuous disruptions to the LFT flow. These disruptions propagate upstream, where the PL controller attempts to adjust the CAVs in the available gaps as they encounter hindrances caused by HDVs, leading to increased lateral movements. At vehicle densities lower than 100 veh/km (as shown in Figure~\ref{fig: relative flow and speed}), these hindrances only cause an increase in lateral movements without significantly reducing LFT flow, as there is sufficient space available for the PL controller to maneuver the CAVs through the HDVs. However, at higher vehicle densities and HDV penetrations, many CAVs are hindered by the HDVs, with CAVs adjusting their speeds and acceleration according to the HDVs. This leads to the formation of traffic waves, as shown in Figure~\ref{fig: space time}, causing a significant drop in performance.

\begin{figure}[!tb]
     \centering
     \begin{subfigure}[b]{1\textwidth}
         \centering
         \includegraphics[width=\textwidth,trim={0cm 19.5cm 0cm 19.5cm},clip]{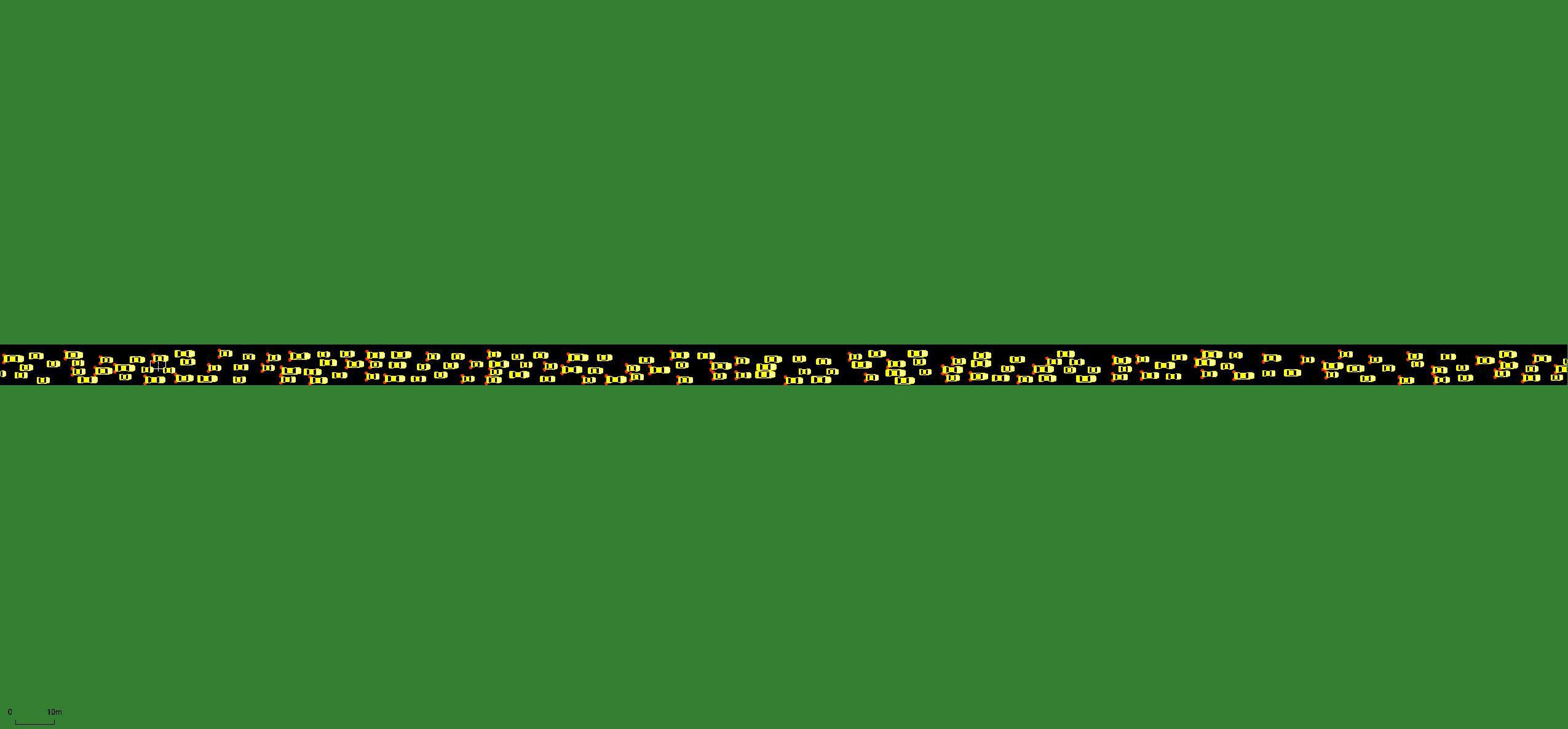}
         \caption{Initial positions at $t=0$ min.}
         \label{fig: SUMO Pics (a)}
     \end{subfigure}
     \hfill
     \begin{subfigure}[b]{1\textwidth}
         \centering
        \includegraphics[width=\textwidth,trim={0cm 19.5cm 0cm 19.5cm},clip]{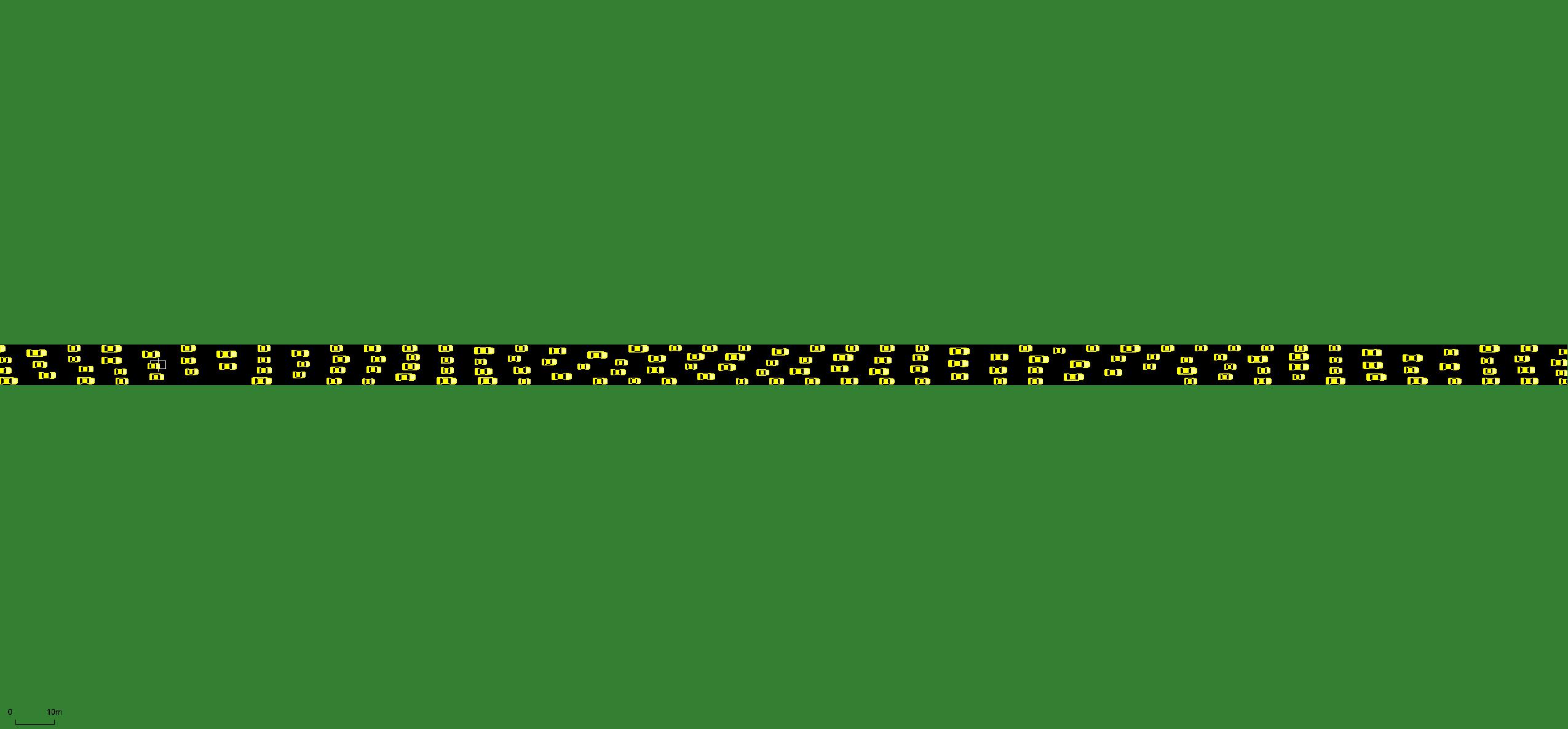}
         \caption{All-CAVs scenario at $t=50$} min.
        \label{fig: SUMO Pics (b)}
     \end{subfigure}
     \hfill
     \begin{subfigure}[b]{1\textwidth}
         \centering
        \includegraphics[width=\textwidth,trim={0cm 19.5cm 0cm 19.5cm},clip]{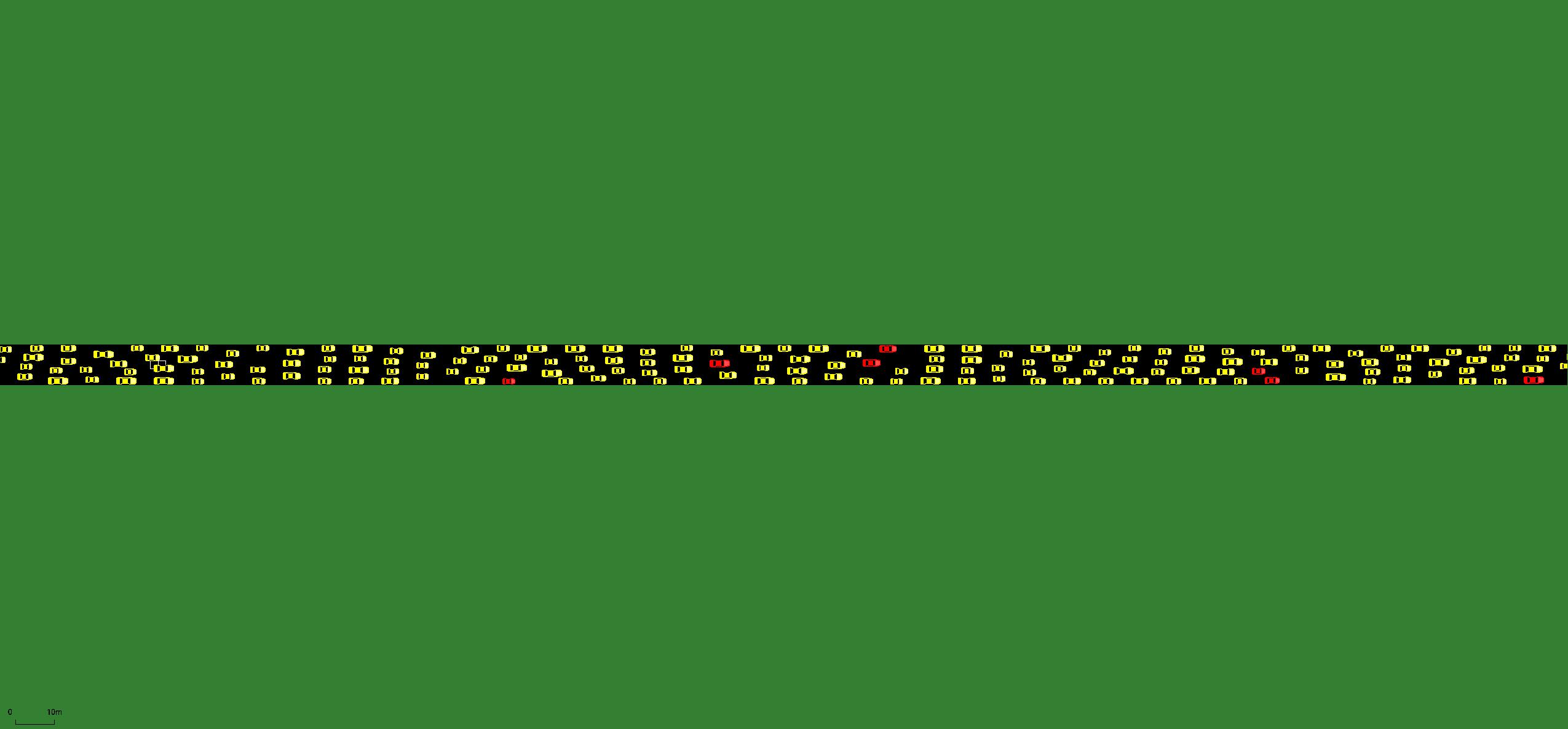}
         \caption{5\% HDVs at $t=50$ min.}
                  \label{fig: SUMO Pics (c)}
     \end{subfigure}
     \hfill
    \begin{subfigure}[b]{1\textwidth}
         \centering
        \includegraphics[width=\textwidth,trim={0cm 19.5cm 0cm 19.5cm},clip]{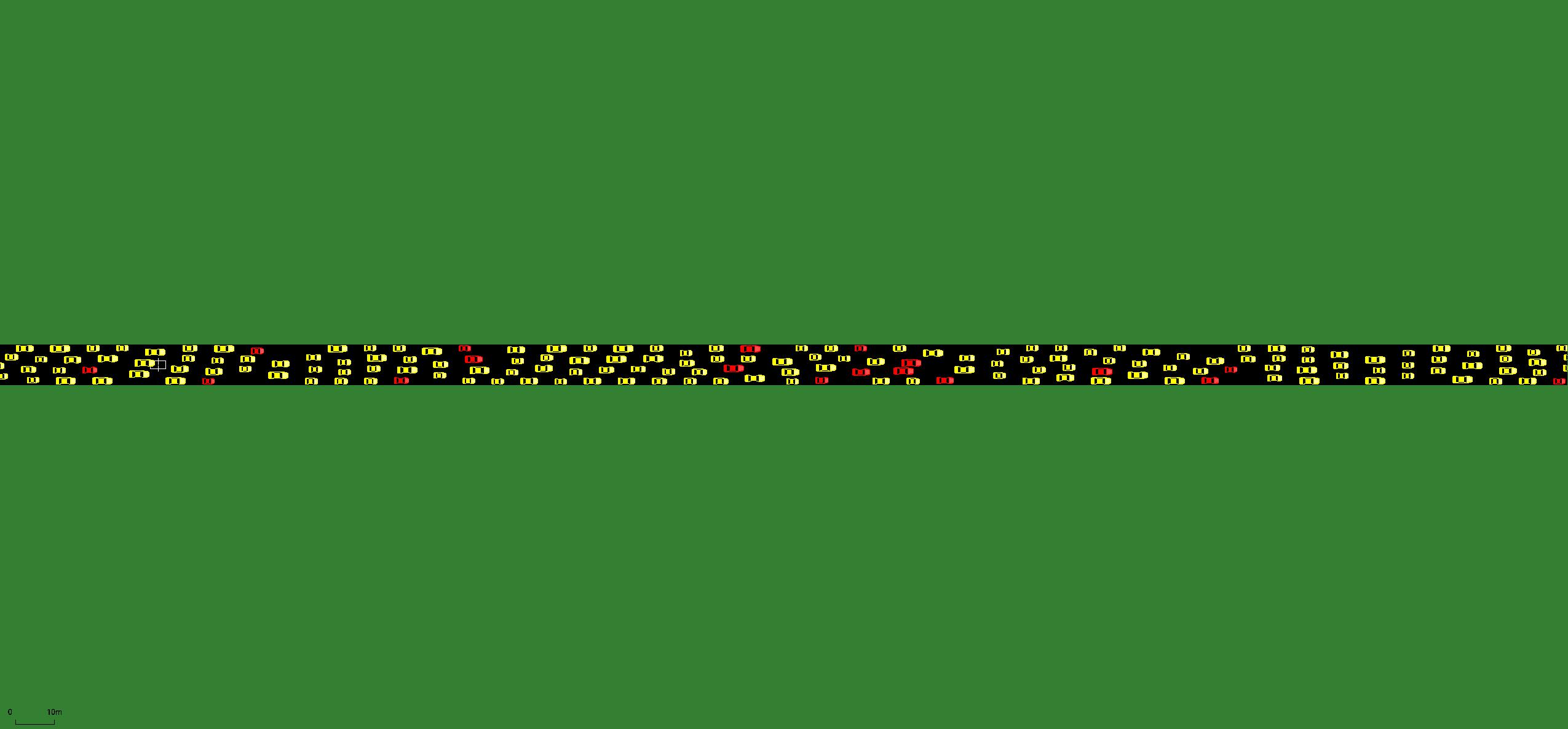}
         \caption{10\% HDVs at $t=50$ min.}
                  \label{fig: SUMO Pics (d)}
     \end{subfigure}
     \hfill
    \begin{subfigure}[b]{1\textwidth}
         \centering
        \includegraphics[width=\textwidth,trim={0cm 19.5cm 0cm 19.5cm},clip]{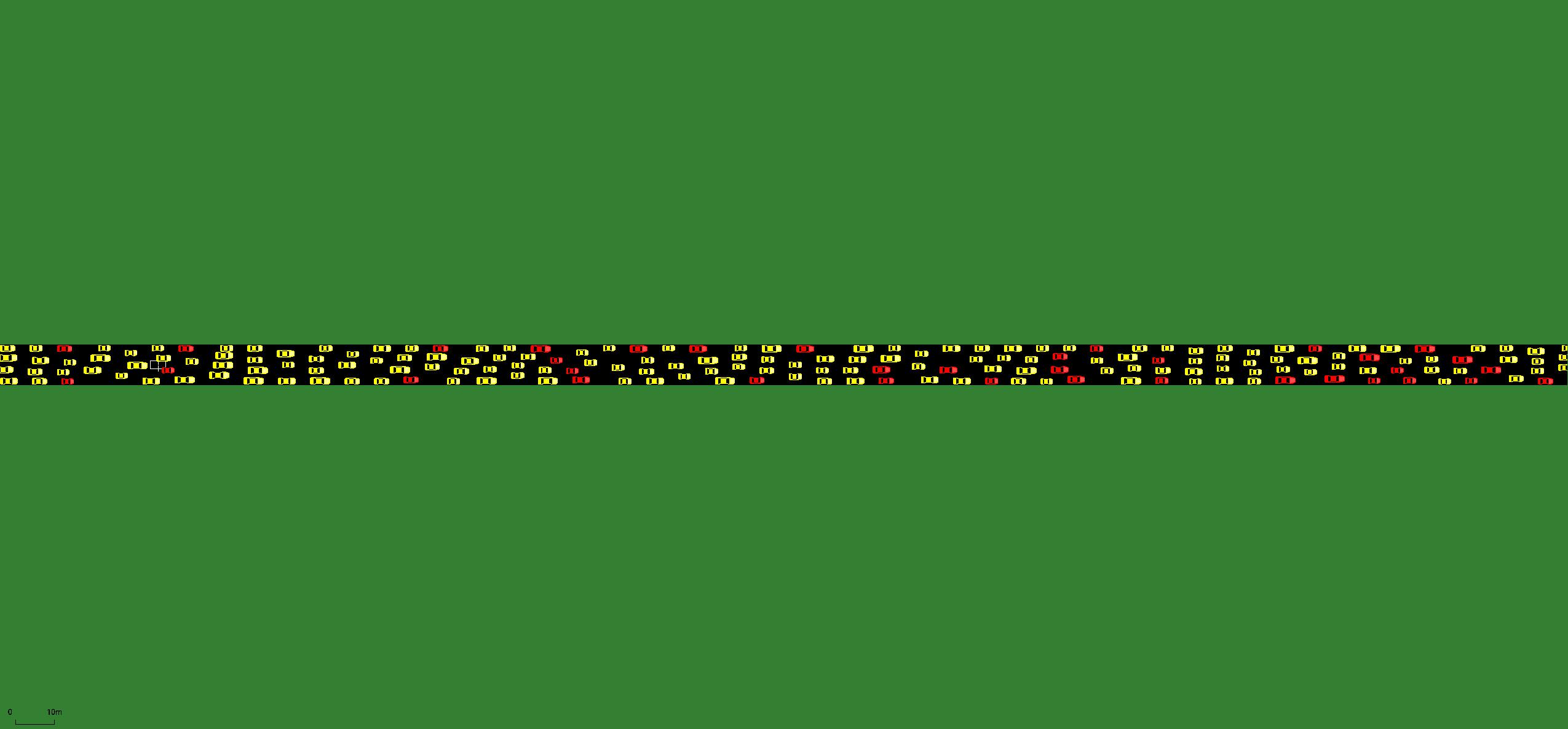}
         \caption{20\% HDVs at $t=50$ min.}
                  \label{fig: SUMO Pics (d_2)}
     \end{subfigure}
    \hfill
    \begin{subfigure}[b]{1\textwidth}
         \centering
        \includegraphics[width=\textwidth,trim={0cm 19.5cm 0cm 19.5cm},clip]{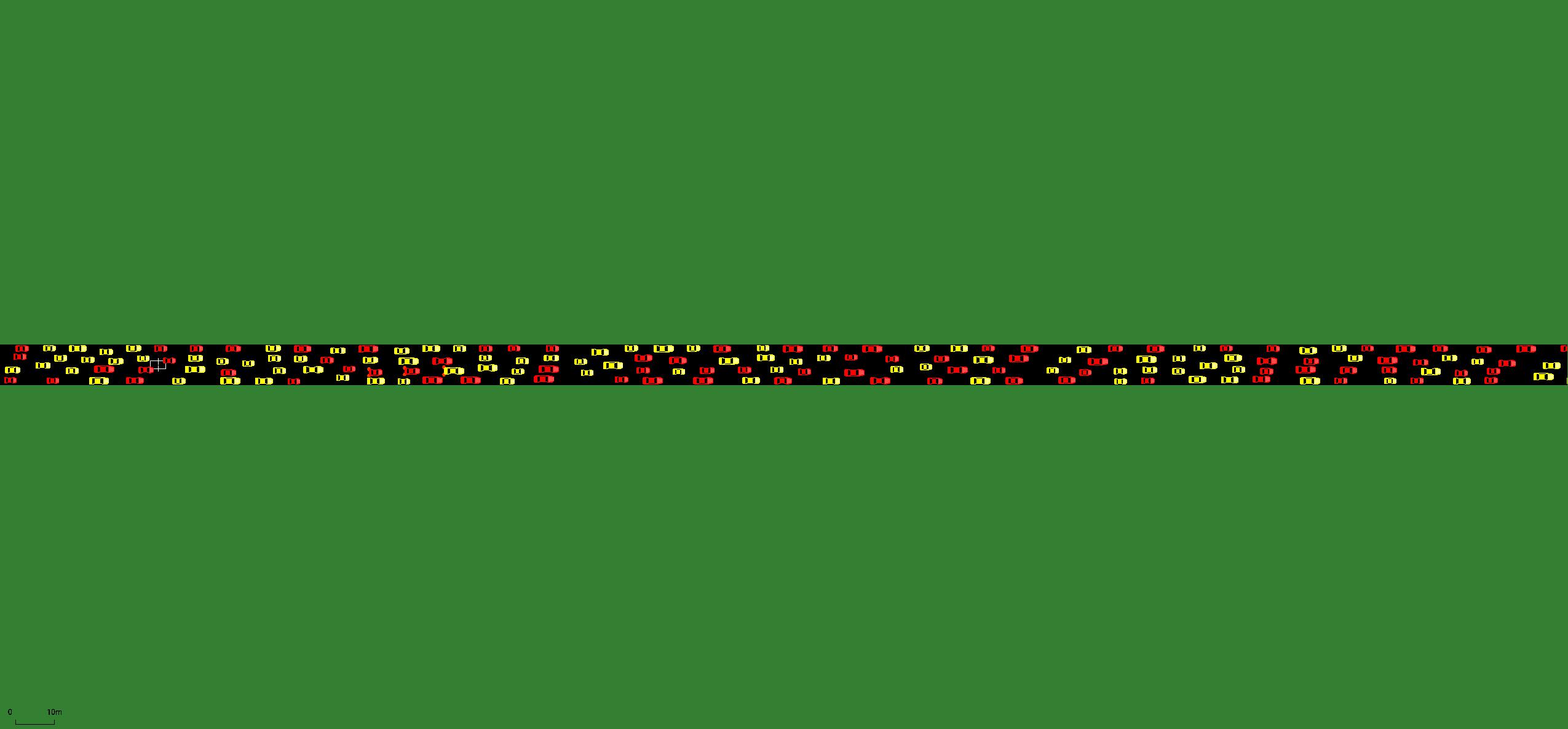}
         \caption{50\% HDVs at $t=50$ min.}
                  \label{fig: SUMO Pics (e)}
     \end{subfigure}
    \hfill
    \begin{subfigure}[b]{1\textwidth}
         \centering
        \includegraphics[width=\textwidth,trim={0cm 19.5cm 0cm 19.5cm},clip]{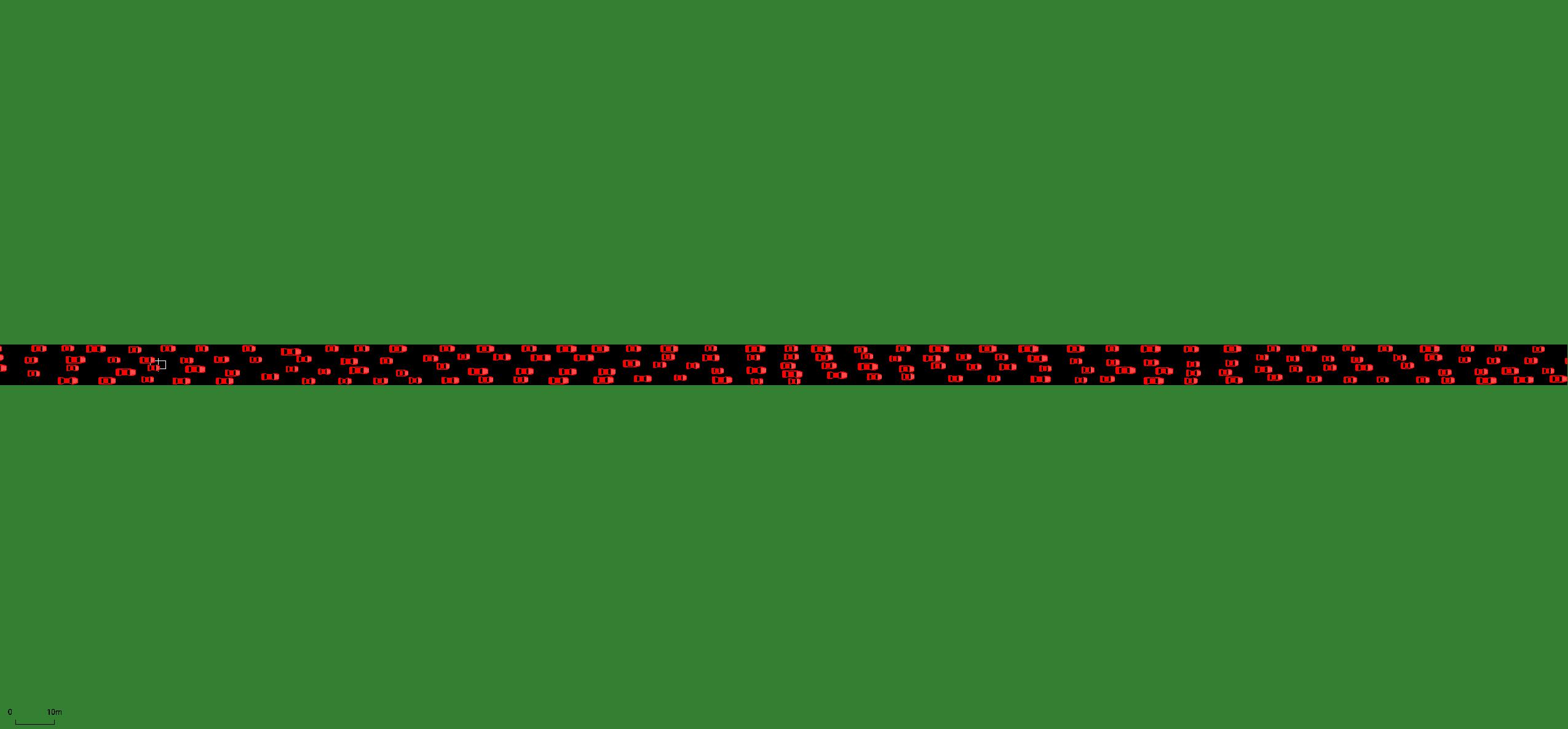}
         \caption{100\% HDVs at $t=50$ min.}
                  \label{fig: SUMO Pics (f)}
     \end{subfigure}
        \caption{Examples of traffic situations with vehicle density of 400~$veh/km$ and PL controller. The starting 400~m of the ring road is shown. The yellow and red vehicles represent CAVs and HDVs, respectively. The initial positions for scenarios with HDVs remain the same as with all-CAVs scenario; only the appropriate proportion of vehicles are set as HDVs.}
        \label{fig: SUMO Pics}
\end{figure}

Figure~\ref{fig: relative flow and speed} also shows that the lateral movements continue to increase with higher penetration rates of HDVs till a saturation point is reached, after which they begin to decline again. As the vehicle densities increase, the saturation point occurs at a lower penetration rate of HDVs. This phenomenon can be explained as follows: at lower penetration rates, the upstream CAVs can better adjust their positions in front of the hindrances caused by HDVs since the CAVs can influence other CAVs, leading to better utilization of space and higher lateral movement; however, as the proportion of HDVs increases, CAVs are unable to do that, leading to lower lateral movement and higher gaps between vehicles. This is visible by comparing Figure~\ref{fig: SUMO Pics (c)}, Figure~\ref{fig: SUMO Pics (d)}, where the portions of road with tightly packed CAVs are larger than scenarios with 5\% HDVs scenario. This happens till the saturation point, after which the proportion of HDVs is so high that the CAVs find limited opportunity to exert artificial forces, and consequently, the lateral movement and the efficient road space utilization decreases, as illustrated in Figure~\ref{fig: SUMO Pics (e)}. This saturation point happens earlier for higher densities because the number of HDVs needed to cause this lateral movement decline is achieved earlier for higher vehicle densities.

\begin{figure}[!tb]
     \centering
     \begin{subfigure}[b]{1\textwidth}
         \centering
         \includegraphics[width=\textwidth]{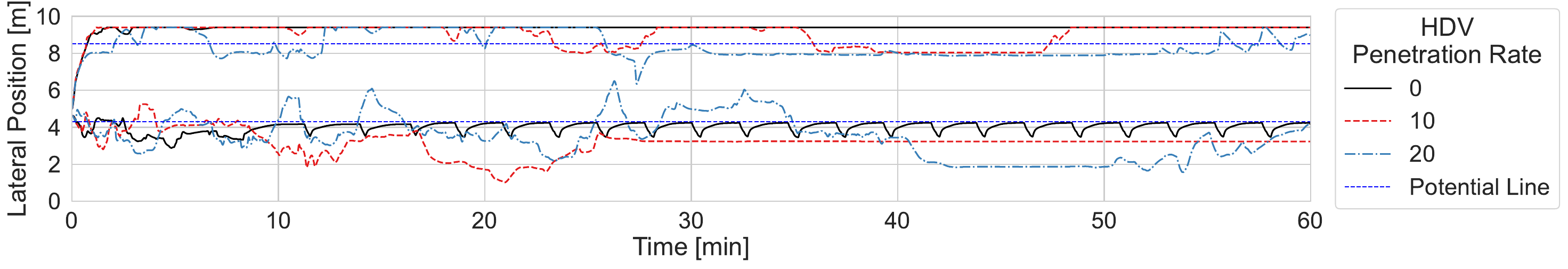}
         \caption{200~$veh/km$}
         \label{fig: CAV trajectory (a)}
     \end{subfigure}
     \hfill
     \begin{subfigure}[b]{1\textwidth}
         \centering
         \includegraphics[width=\textwidth]{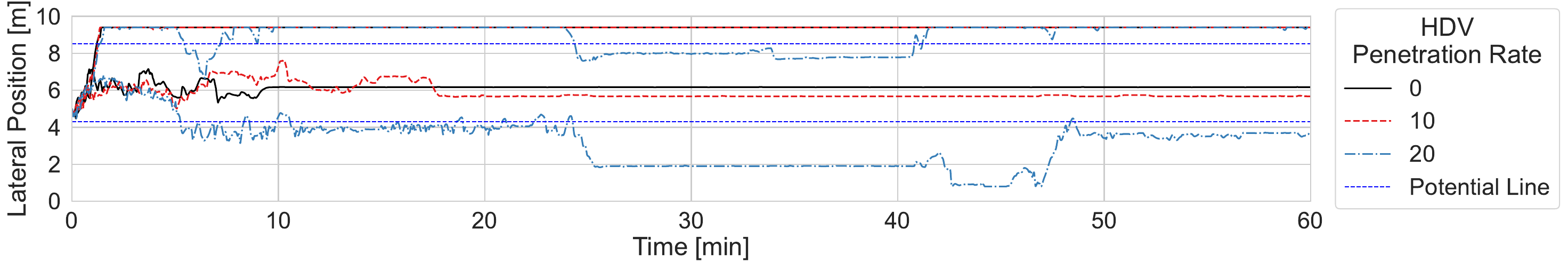}
         \caption{400~$veh/km$}
         \label{fig: CAV trajectory (b)}
     \end{subfigure}
        \caption{The lateral positions of two CAVs. The inclusion of HDVs causes significant changes in the lateral movement of CAVs.}
        \label{fig: CAV trajectory}
\end{figure}

The above phenomena can also be observed from the development of lateral positions of CAVs over time, as shown in Figure~\ref{fig: CAV trajectory}. The behavior of the PL controller with all-CAVs scenario is discussed first. The PL controller tries to minimize the lateral movement by assigning PL according to desired speeds. At the beginning of the simulation, there are usually higher lateral movements since all CAVs try to reach their assigned PLs. Over time, the CAVs settle down at a lateral position and the lateral movements are minimized subsequently. It is possible that these settling points are not exactly on the assigned PL and are slightly shifted, according to the location of the PL and the artificial forces from surrounding vehicles. It is observed that this deviation of the settling point is larger near the road boundaries, as shown by all-CAVs scenarios in Figure~\ref{fig: CAV trajectory}. The CAVs with assigned PLs near the boundaries experience artificial forces from the surrounding vehicles from one side only, and in the absence of any artificial forces to counter them, these CAVs are pushed further towards the boundaries. In contrast, the CAVs with assigned PLs near the center, experience artificial forces from both sides, making it easier for them to stay on the assigned PL. However, at higher vehicle densities, even the CAVs at the center may have significant deviation due to the CAVs not finding enough space and opportunity to remain on the assigned PL, as shown by the all-CAVs scenario in Figure~\ref{fig: CAV trajectory (b)}. 

As the penetration of HDVs increases, Figure~\ref{fig: CAV trajectory} shows that the lateral movement of CAVs is significantly impacted. The main reason for this is that the HDVs do not target aligning themselves with specific PLs. This causes them to travel without coordination with CAVs, resulting in uncoordinated LFT forces on CAVs and significantly higher lateral movements. This also results in CAVs experiencing long episodes of going far away from the assigned PL. This phenomenon is observed to be more severe for higher vehicle densities; however, it is not limited to higher vehicle densities and can be observed even for small vehicle densities. Nonetheless, unlike at higher vehicle densities, at lower vehicle densities, this does not significantly affect traffic flow due to having sufficient space. Overall, this shows that even a small proportion of HDVs has the potential to not only degrade LFT flow but also significantly affect the characteristic features of certain LFT controllers, for example, the feature of reduced lateral movements in the case of PL controller.

\subsection{Improvements using Adaptive Potential Lines Controller}

\begin{figure}[tb]
     \centering
     \includegraphics[width=1\textwidth]{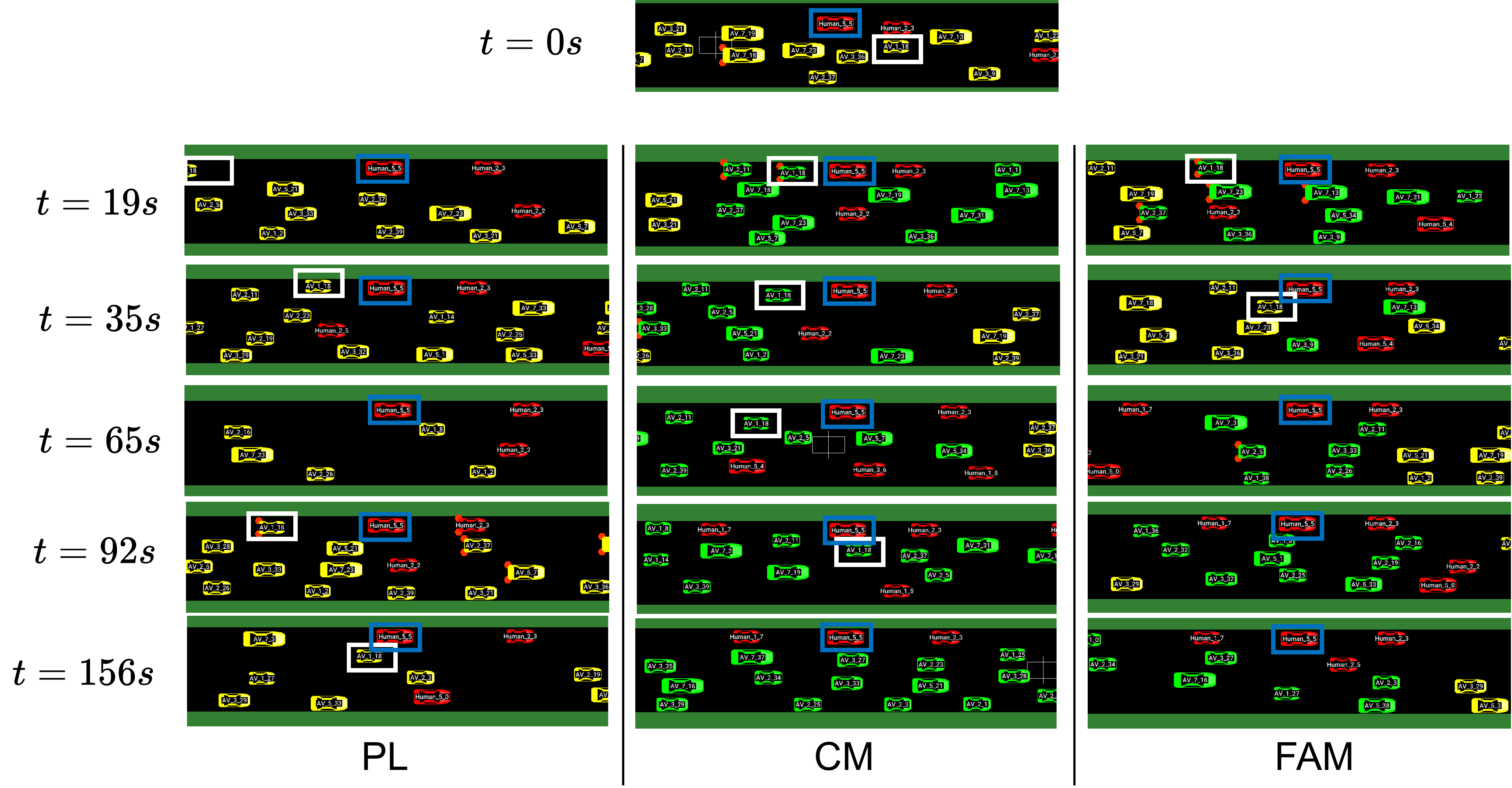}
     \caption{Comparison of the temporal development of PL and two APL controllers for 250~$veh/km$ and 20\% HDVs. The red and yellow vehicles represent HDVs and CAVs, respectively. The green vehicles represent the CAVs under the influence of APL corridors. The blue and white rectangles show the HDV and the CAV focused on in the discussion, respectively. The $X_{CM}$ and $X_{AM}$ are set to 20~m in this example.}
     \label{fig: temporal development APL}
\end{figure}

\begin{figure}[!tb]
     \centering
     \begin{subfigure}[b]{1\textwidth}
         \centering
         \includegraphics[width=\textwidth,trim={10cm 19.5cm 10cm 19.5cm},clip]{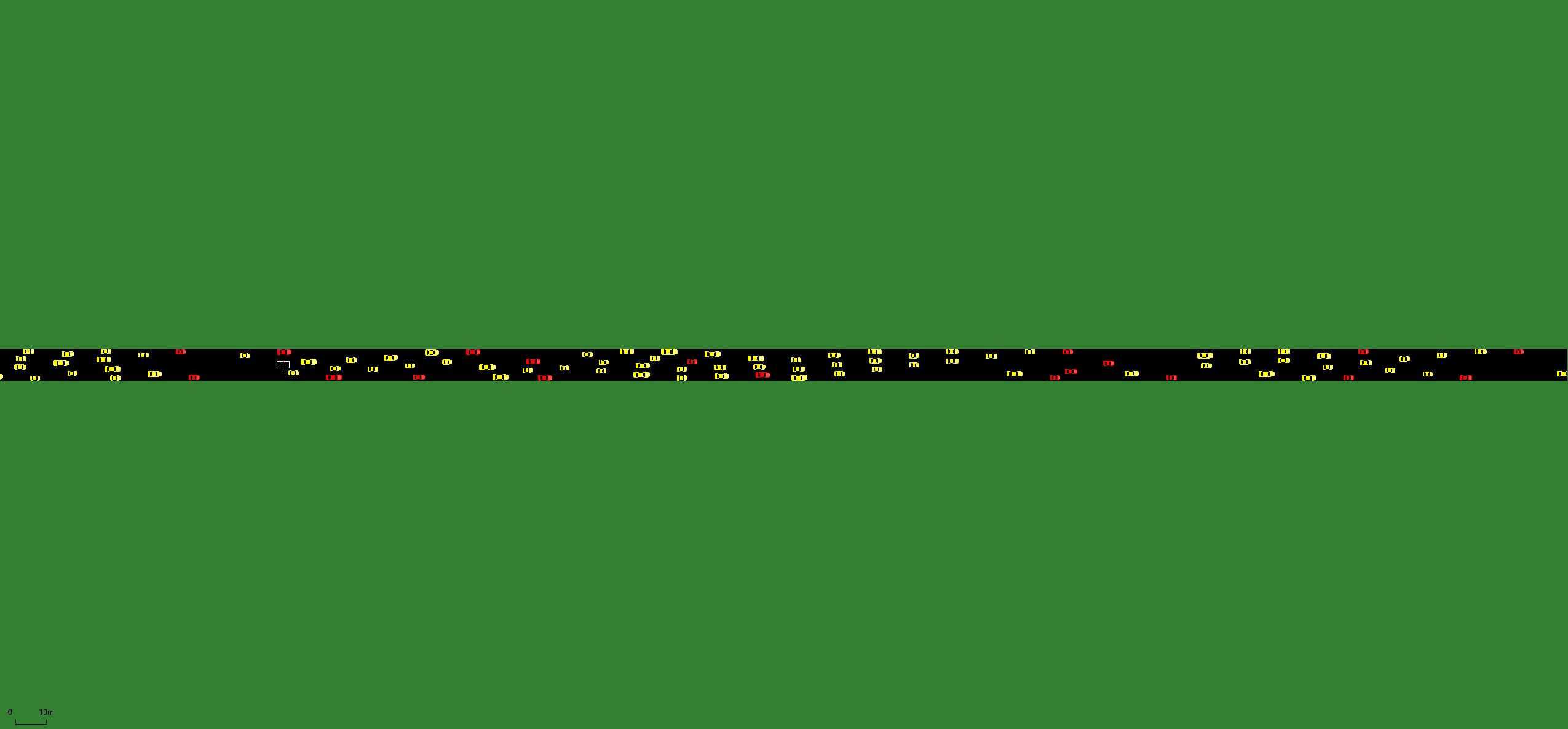}
         \caption{PL}
         \label{fig: SUMO Pics Corridor (a)}
     \end{subfigure}
     \hfill
     \begin{subfigure}[b]{1\textwidth}
         \centering
         \includegraphics[width=\textwidth,trim={10cm 19.5cm 10cm 19.5cm},clip]{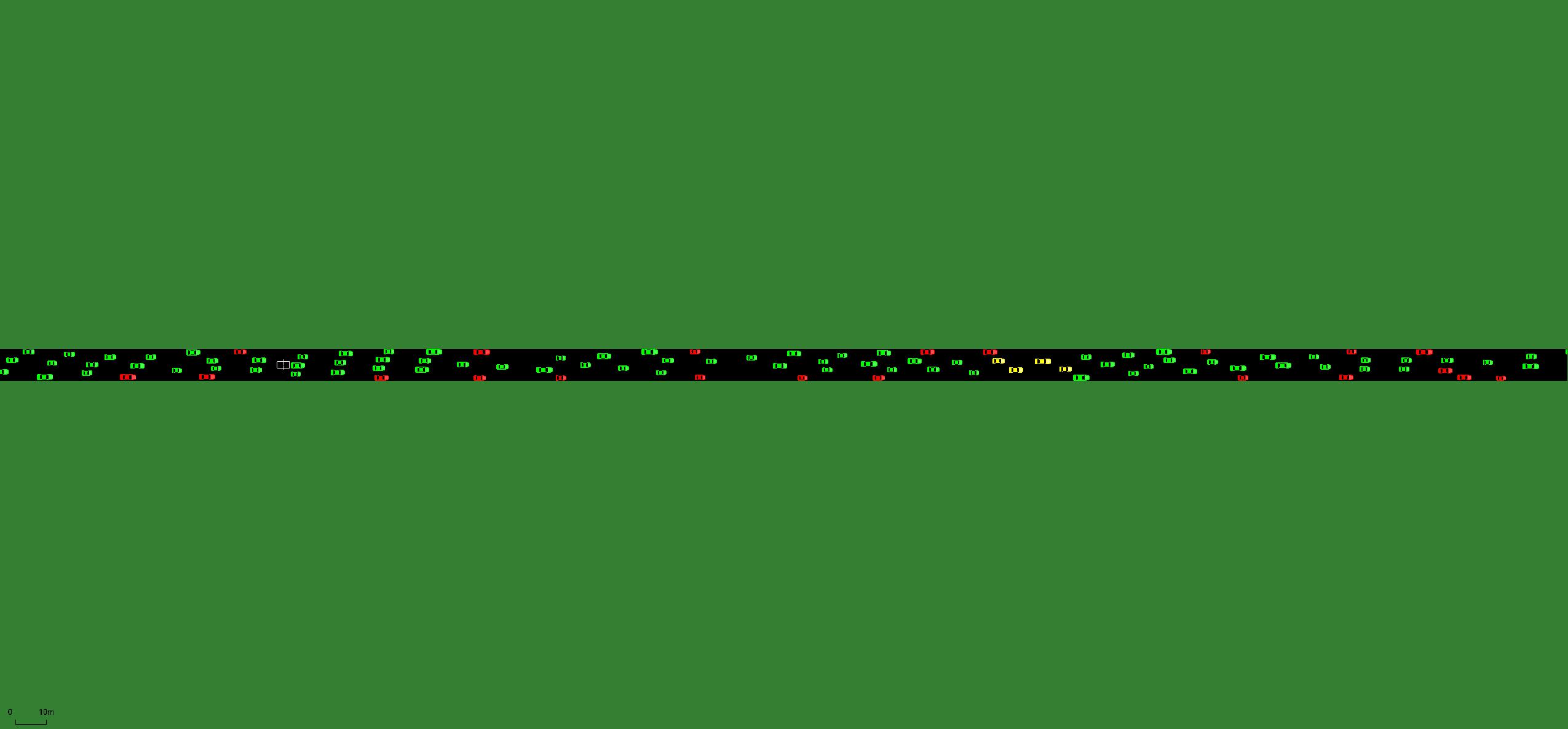}
         \caption{CM}
        \label{fig: SUMO Pics Corridor (b)}
     \end{subfigure}
     \hfill
     \begin{subfigure}[b]{1\textwidth}
         \centering
         \includegraphics[width=\textwidth,trim={10cm 19.5cm 10cm 19.5cm},clip]{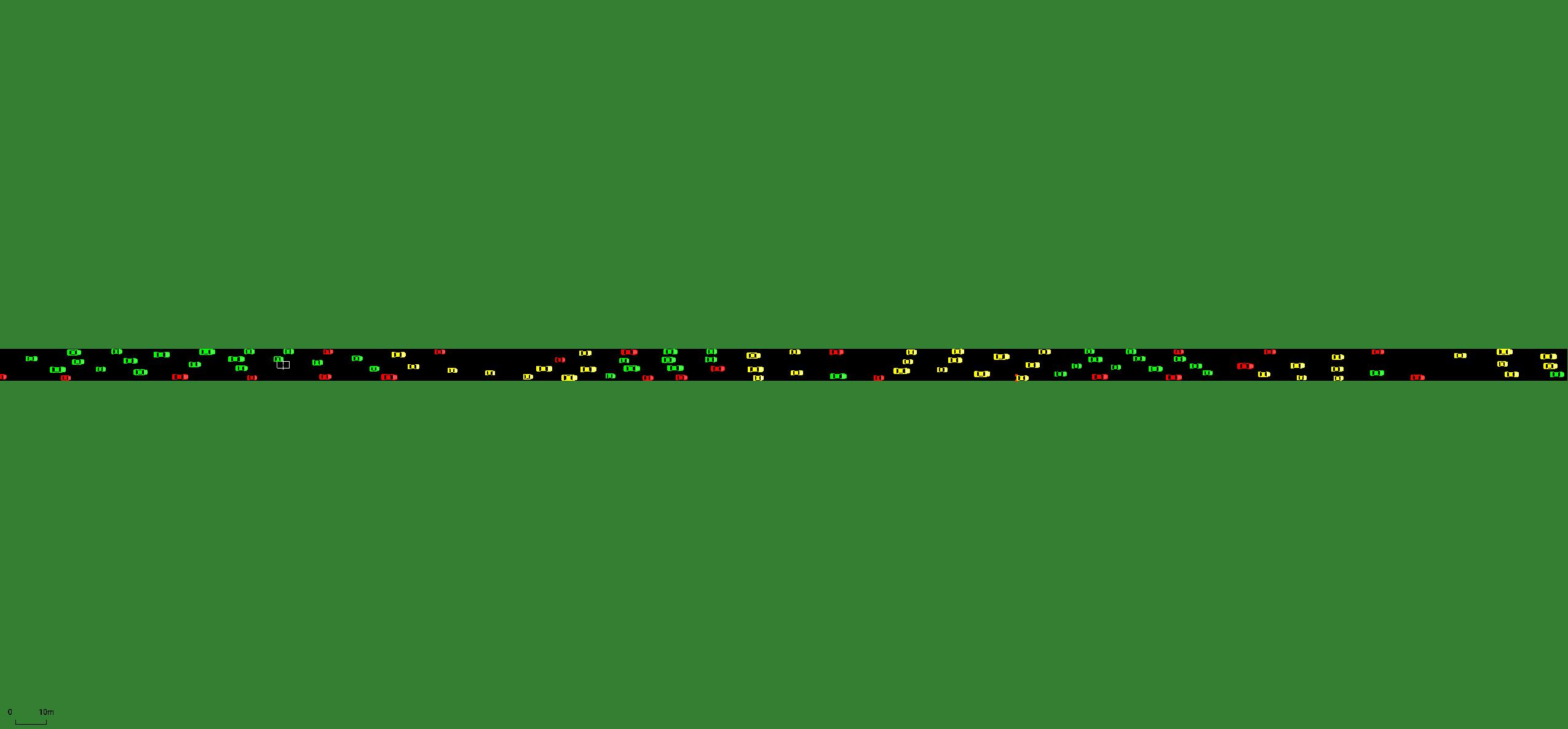}
         \caption{FAM}
                  \label{fig: SUMO Pics Corridor (c)}
     \end{subfigure}
        \caption{The formation of APL corridors between HDVs for CM and FAM strategies. The CAVs following APL corridors are represented with green color. The figure represents the situation at $t=50$ min with 200~$veh/km$ and 20\% HDVs.}
        \label{fig: SUMO Pics Corridor}
\end{figure}

The results in the previous section highlighted the significant performance drop observed with the penetration of HDVs into LFT. This section presents how much the APL controller introduced can improve this behavior. While the current study focused on the overtaking maneuver of the follower CAVs to develop APL activation procedures, the interaction of multiple CAVs in LFT can have significant impacts. To better understand these interactions in LFT, Figure~\ref{fig: temporal development APL} compares the initial temporal development of PL, CM, and FAM strategies as a CAV (marked with a white rectangle) overtakes an HDV (marked with a blue rectangle). The marked CAV has a high desired speed; thus, the assigned PL is typically on the left side of the road. The CAV starts ($t=0$) in the middle but ends up behind the marked HDV by aligning itself to the assigned PL ($t=19s$). With a simple PL controller, the CAV remains behind the HDV. The gap between the two increases ($t=65s$) and decreases ($t=92s$) with the acceleration and deceleration of the HDV, contributing to the formation of traffic waves. The marked CAV overtakes the HDV when the following CAV exerts lateral force, allowing the marked CAV to change its lateral position and overtake the HDV ($t=156s$). Nonetheless, such an overtake is purely dependent on the developing situation, and the PL controller is not taking active measures to avoid the adverse effects of HDVs.

In contrast, the APL controllers allow the marked CAV to overtake the HDV significantly earlier. However, the specific details depend on the APL strategy used. For example, since CM applies APL within a fixed distance $X_{CM}$ behind each HDV, the APL corridor extends to multiple CAVs (including the ones behind the marked CAV), as shown at $t=19s$. Consequently, many CAVs try to reach the newly assigned PLs. The new PLs would be more accessible for CAVs laterally close to modified PLs ($t=35s)$. Thus, these CAVs can accelerate more quickly than the marked CAV since HDVs do not block them in the APL corridor. In turn, they nudge the marked CAV away as it tries to overtake the HDV, as shown at $t=65s$ for CM. However, as soon as sufficient space is available, the marked CAV and the one behind it change their lateral position and overtake the HDV at $t=92s$. Figure~\ref{fig: temporal development APL} presents FAM's temporal development as the other example of APL controller. Compared to the CM, the FAM strategy limits the APL corridor up to the marked CAV ($t=19s$) till it changes its lateral position and the corridor is extended to the newer follower CAV; thus, fewer CAVs are competing with the marked CAV, or they compete with some delay when the APL corridor is formed due to the newer follower CAV. This provides enough time and space for the marked CAV to change its lateral position and overtake the HDV earlier than the CM strategy. It is also worth noting that even when the modified PL is deactivated in FAM for some time ($t=35s$) due to the condition of the average surrounding speed, the marked CAV does not go back behind the HDV; instead, it continues to overtake due to the artificial forces exerted by the new follower CAV of HDV. This shows that the problem of developing LFT strategies with a mixture of HDVs is not merely a matter of how a blocked CAV overtakes HDVs; instead, it has multiple facets regarding how multiple CAVs would interact to accommodate HDVs while improving the overall situation. 

For the APL strategies, $t=92s$ and $t=156s$ show another phenomenon. Since the longitudinally overlapping regions are combined into single APL corridors, it is observed that many HDVs start to follow each other, forming longitudinal groups of HDVs. This is also visible in Figure~\ref{fig: SUMO Pics Corridor} for an extended road portion, where compared to PL case, HDVs are less dispersed in CM and FAM case. Figure~\ref{fig: SUMO Pics Corridor} also shows that since CM (and NSCM) use a fixed distance ($X_{CM}$) behind each HDV to form APL corridors, it can create long stretches of APL corridors. In contrast, FAM (and SVAM) create shorter APL corridors. Consequently, it can be said that while the adaptive margin-based strategies favor the overtaking maneuvers of the follower CAVs, the constant margin-based strategies favor the movement of a larger group of CAVs using longer APL corridors.

\begin{figure}[tb]
     \centering
     \includegraphics[width=0.9\textwidth]{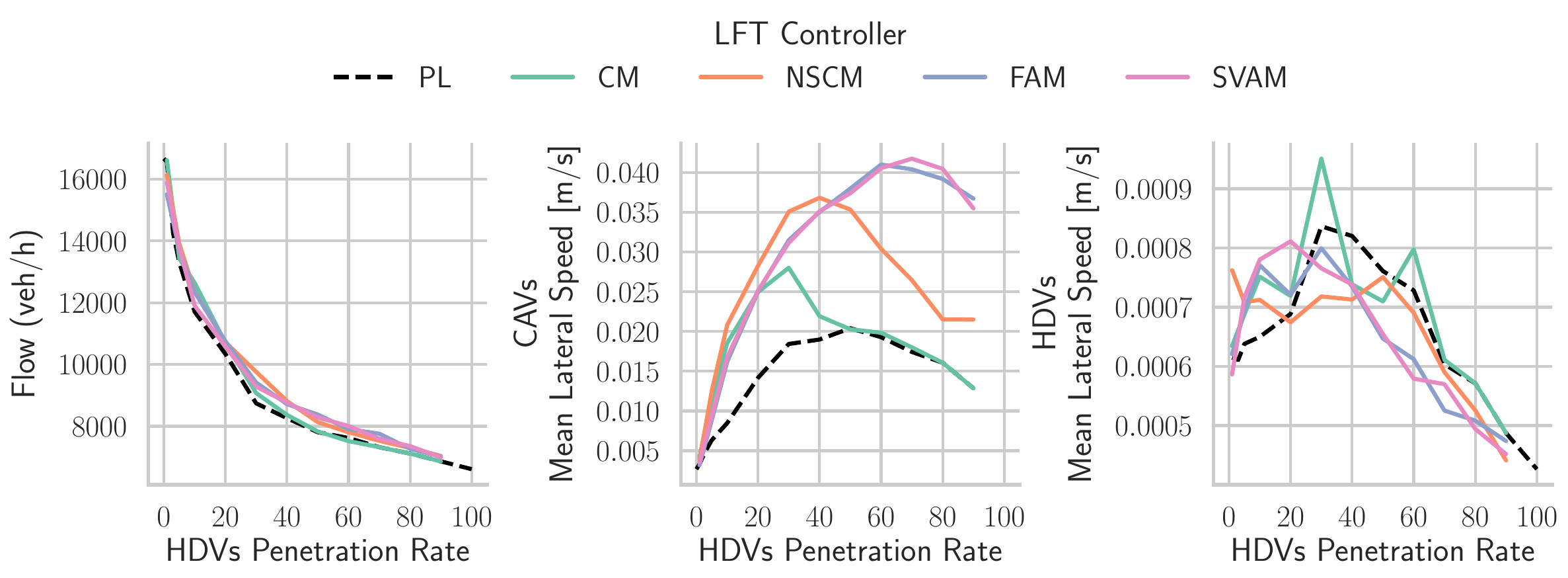}
     \caption{The traffic flow rates and mean lateral speeds for different APL strategies and HDV penetrations. The vehicle density is set to 200~$veh/km$}.
     \label{fig: adaptive pl hdv overall}
\end{figure}

% \begin{figure}[tb]
%      \centering
%      \includegraphics[width=1\textwidth]{figures/trc_second_revision_new/nudge_1_param_40_adaptive_pl_flow_diff.pdf}
%      \caption{The difference of traffic flow values of APL and PL controllers under the same settings.}
%      \label{fig: adaptive pl hdv penetration}
% \end{figure}

\begin{figure}[!tb]
     \centering
     \begin{subfigure}[b]{1\textwidth}
         \centering
         \includegraphics[width=\textwidth]{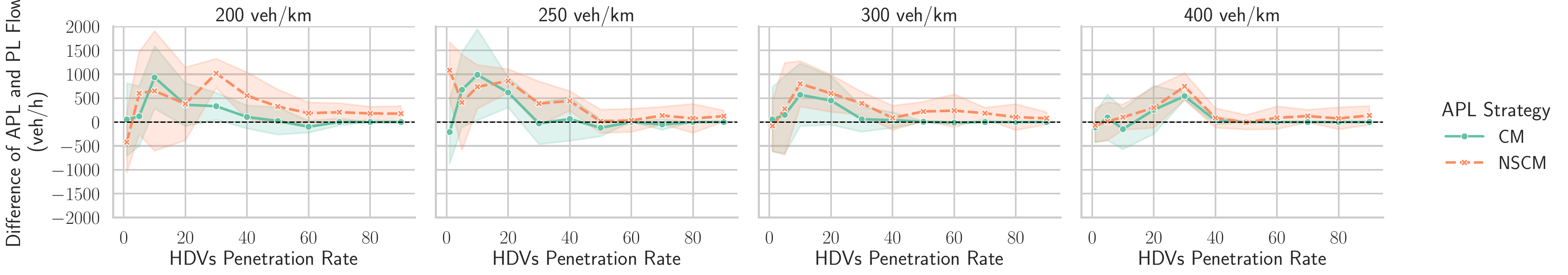}
         \caption{Constant margin based APL strategies}
         \label{fig: adaptive pl hdv penetration (a)}
     \end{subfigure}
     \hfill
     \begin{subfigure}[b]{1\textwidth}
         \centering
         \includegraphics[width=\textwidth]{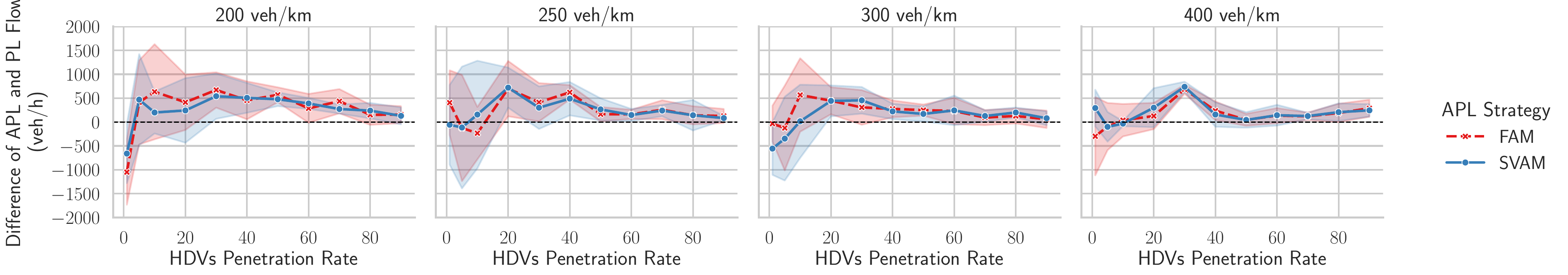}
         \caption{Adaptive margin based APL strategies}
         \label{fig: adaptive pl hdv penetration (b)}
     \end{subfigure}
        \caption{The difference in traffic flow values of APL and PL controllers under the same settings. The shaded areas show the standard deviation.}
        \label{fig: adaptive pl hdv penetration}
\end{figure}

Figure\ref{fig: adaptive pl hdv overall} compares the overall flow and the average lateral speeds of APL and PL controllers for a density of 200~$veh/km$. The first thing to note is the general trend of a significant decrease in LFT traffic flow with HDV penetration, even with the APL controller. Nonetheless, the APL controller significantly improves over the PL controller, especially in the range of 5-60\% HDVs penetration. To compare this further, Figure~\ref{fig: adaptive pl hdv penetration} shows the difference in traffic flow values for APL and PL controllers. For each APL strategy, the improvement generally peaks at a certain HDV penetration and then declines. The CM and NSCM provide the best improvements over PL for an HDV penetration lower than 40\%, causing a peak average traffic flow of around 1000~$veh/h$ (8-12\% improvement) with a density of 200-250~$veh/km$. It should be noted that the performance improvement depends significantly on the initial traffic situation (i.e., simulation seed) and the LFT parameter values used. For example, an additional traffic flow of 2500~$veh/h$ (24\% improvement) was observed for one of the seeds with the NSCM method, 10\% HDV penetration, and 200~$veh/km$.

\begin{figure}[tb]
     \centering
     \includegraphics[width=1\textwidth]{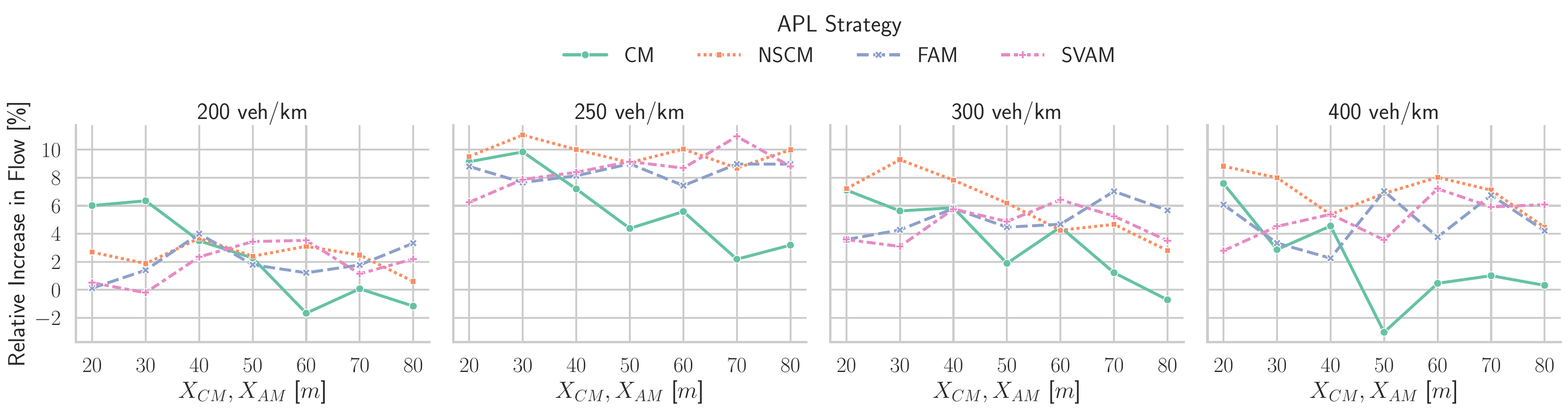}
     \caption{Impact of changing the parameter values for various APL strategies and 20\% HDVs. The y-axis shows the relative increase in the traffic flow compared to the PL controller's flow under the same setting.}
     \label{fig: adaptive pl hdv parameters}
\end{figure}

\begin{figure}[!tb]
     \centering
     \begin{subfigure}[b]{0.49\textwidth}
         \centering
         \includegraphics[width=\textwidth]{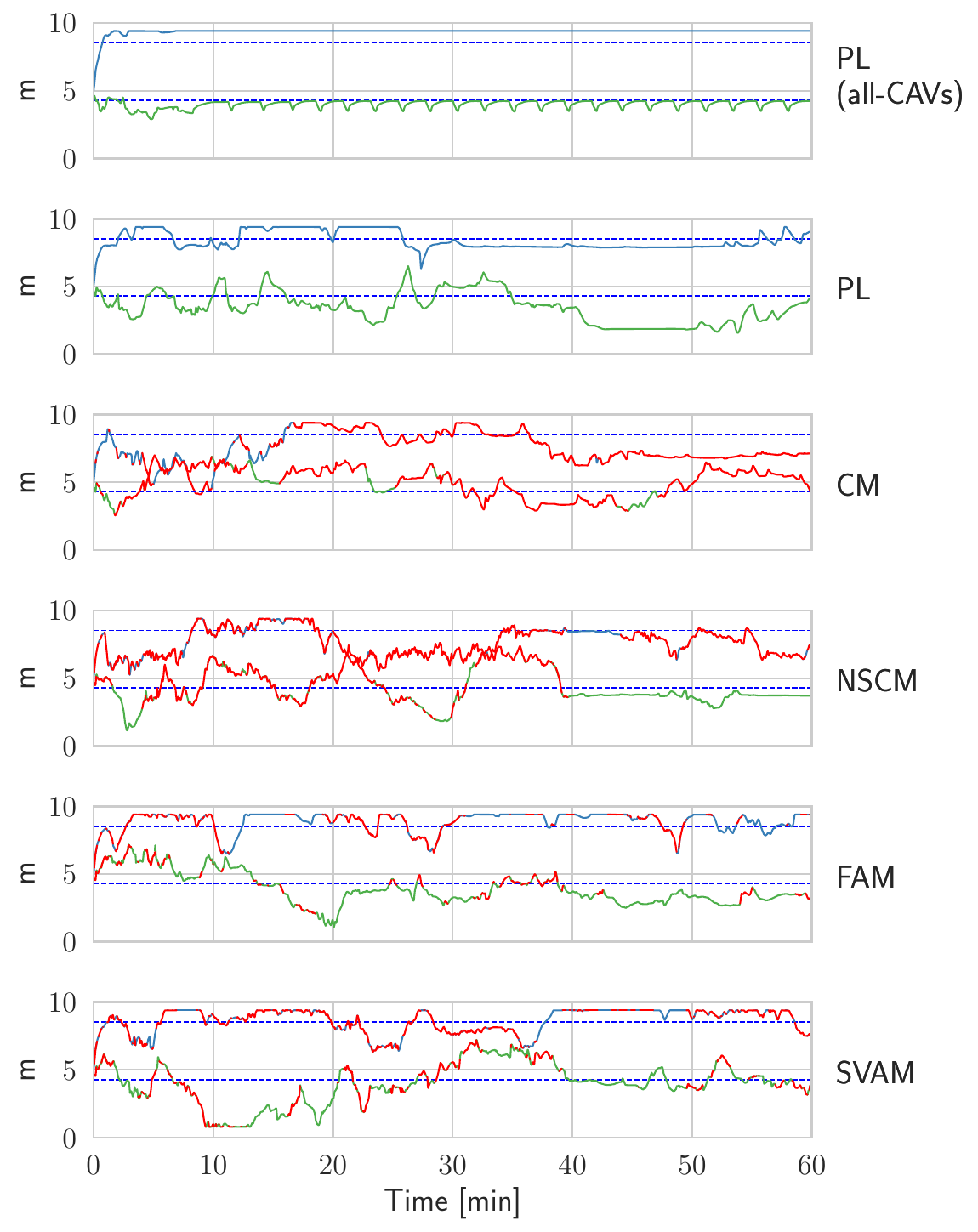}
         \caption{200~$veh/km$}
         \label{fig: APL trajectory (a)}
     \end{subfigure}
     \hfill
     \begin{subfigure}[b]{0.49\textwidth}
         \centering
         \includegraphics[width=\textwidth]{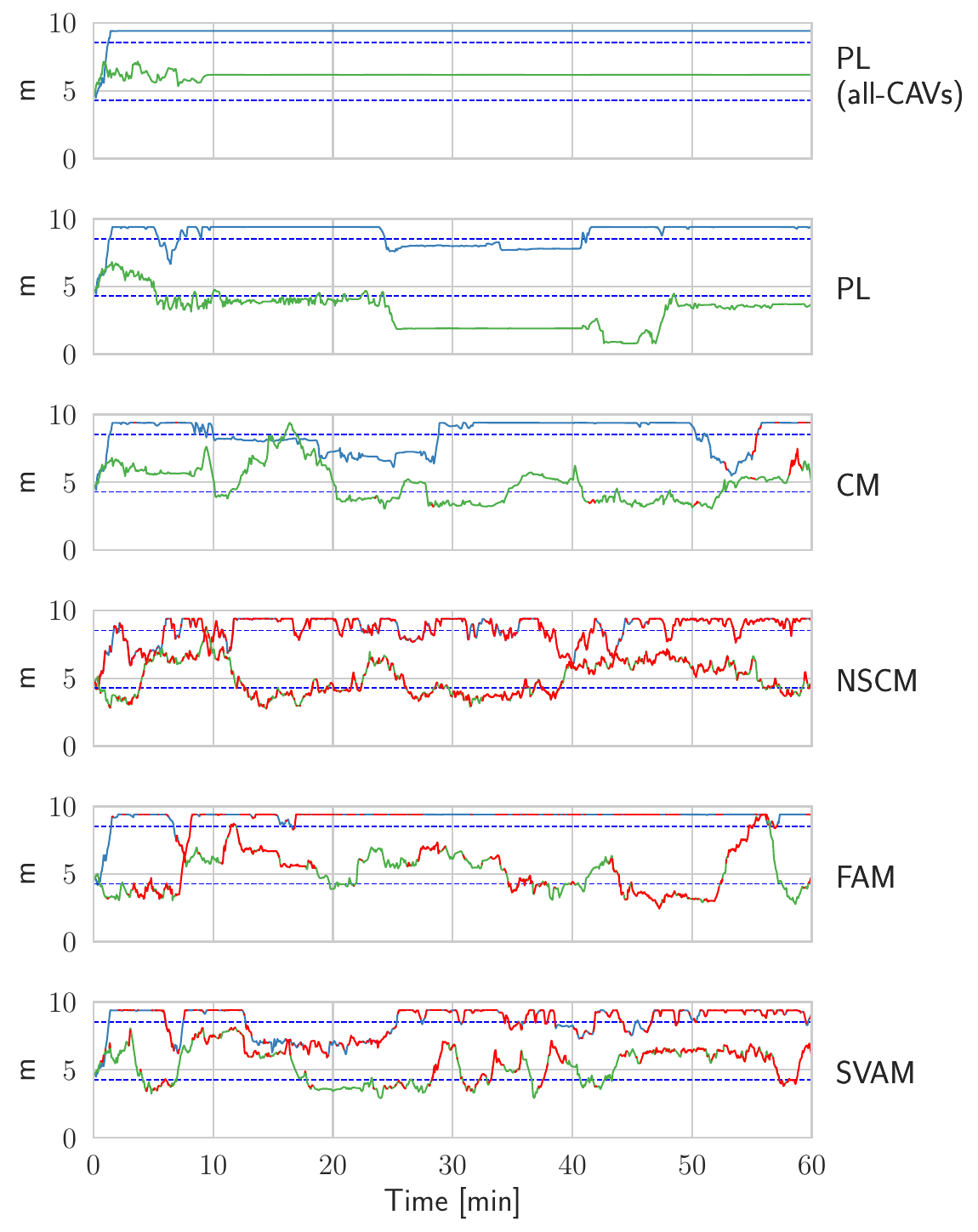}
         \caption{400~$veh/km$}
         \label{fig: APL trajectory (b)}
     \end{subfigure}
        \caption{The lateral positions of a CAV under APL controllers with 20\% HDVs (except all-CAVs case). The red lines indicate the trajectory under the influence of APL corridors.}
        \label{fig: APL trajectory}
\end{figure}

Additionally, the improvements of NSCM are found to be more general regarding HDV penetration than CM. This means that NSCM continues to provide performance improvement for a wider range of HDVs penetration than CM. This is due to the way APL corridors are formed, where both longitudinally overlapping regions are first combined into a single region. Then, the lateral space available between HDVs in each region is used for APL corridors. With increasing HDV penetration, there are longitudinally more overlapping regions and less empty lateral space available for each region to form APL corridor. Consequently, since CM uses all HDVs in this process, it finds it challenging to form APL corridors and shows degraded performance earlier. This reason also applies when higher values of $X_{CM}$ are used with CM, as demonstrated in Figure~\ref{fig: adaptive pl hdv parameters}, or when a higher vehicle density is simulated. Consequently, it is worth noting in Figure~\ref{fig: adaptive pl hdv parameters} that CM's performance is generally improved when a lower $X_{CM}$ is used. In contrast, NSCM only considers an HDV for APL corridors when the HDV is moving slower than the surrounding vehicles. This reduces the overlapping longitudinal regions and allows NSCM to form more APL corridors than CM. This is also visible through the lateral trajectory plots in Figure~\ref{fig: APL trajectory (b)}, where the vehicle experiences APL corridors in NSCM much more than CM.

Figure~\ref{fig: adaptive pl hdv penetration (b)} also shows that the improvements of FAM and SVAM are smaller than CM and NSCM, especially for starting 0-40\% HDV penetration rate. However, FAM and SVAM are observed to be even more general than NSCM strategy in terms of HDV penetration, producing higher improvement than NSCM for scenarios with more than 50\% HDVs. This can be explained by their adaptive nature for forming APL corridors. Since they only consider the distance up to the follower CAV for the APL corridor, both strategies have fewer overlapping regions, resulting in a short but higher number of APL corridors than CM and NSCM. For smaller HDV penetration (0-30\%), such short corridors would be less efficient than long corridors of CM and NSCM since many CAVs would have to laterally adjust their movements whenever they enter or exit such short APL corridors. This is visible in Figure~\ref{fig: APL trajectory (a)}, where CAVs under FAM and SVAM experience shorter episodes of APL corridors than NSCM and CM. This also explains the higher performance deviation for starting 0-30\% HDVs in Figure~\ref{fig: adaptive pl hdv penetration (b)}.

On the other hand, these short APL corridors benefit from high HDV penetration and vehicle density. Under these conditions, the FAM and SVAM can still form APL corridors, unlike CM and NSCM strategies. This allows CAVs to overtake the blocking HDV and accelerate, making these strategies applicable to even high HDV penetration. This is also visible from the higher lateral speed of CAVs in Figure~\ref{fig: adaptive pl hdv overall} for SVAM and FAM strategies than NSCM for HDV penetration of more than 50\%.

Figure~\ref{fig: adaptive pl hdv overall} also shows that the lateral movement of HDVs in the APL controller is significantly different from PL controller. With the formation of APL corridors, the HDVs find more space to change their lateral position to achieve higher speed, leading to increased mean lateral speed for HDVs under APL methods than the PL method. This is especially observed for starting 0-50\% HDV penetration. If these APL corridors are sustained for longer periods, this also leads to an interesting phenomenon, as shown in Figure~\ref{fig: SUMO Pics Corridor (b)}, where HDVs end up following each other and start to form longitudinal groups. For penetration rates higher than 50\%, the proportion of HDVs experiencing more empty space due to the formation of APL corridors is smaller, and thus, the mean lateral speed of the HDVs also decreases, as shown in Figure~\ref{fig: adaptive pl hdv overall}.

\begin{figure}[!tb]
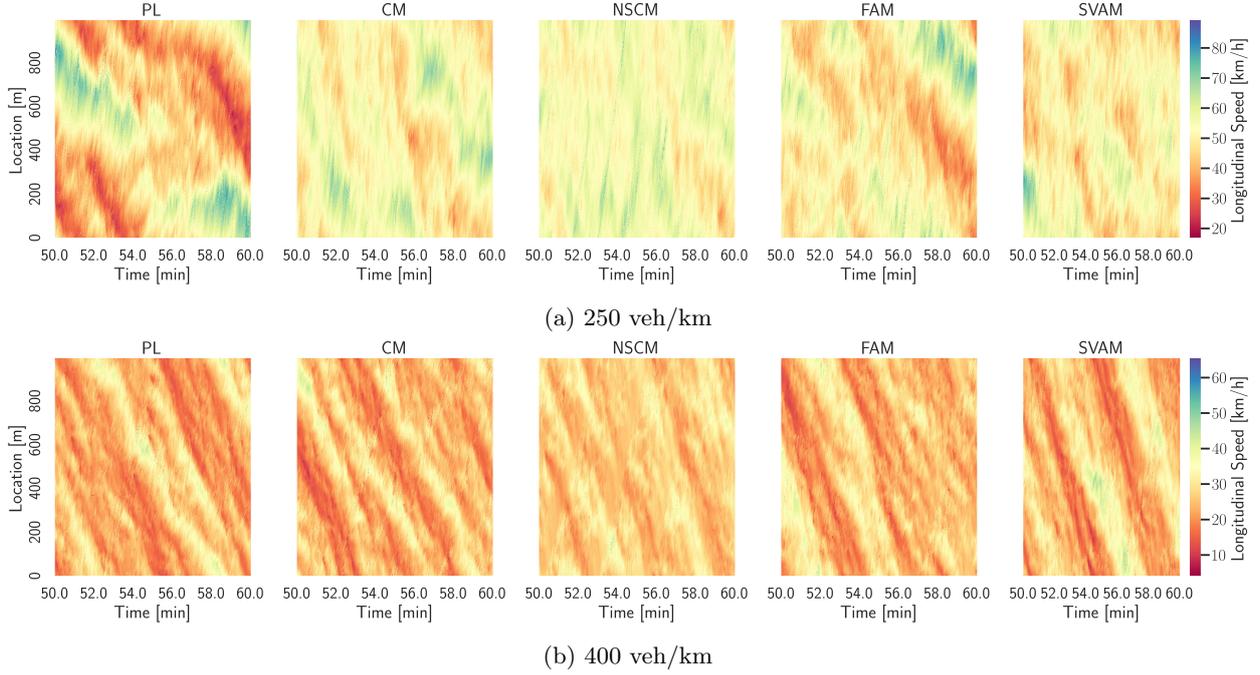

     \centering
     \begin{subfigure}[b]{1\textwidth}
         \centering
         \includegraphics[width=1\textwidth]{figures/trc_second_revision_new/heatmaps/waves_heatmap_adaptive_pl_200_40_seed_2_nudge_1.jpg}
         \caption{200~veh/km}
         \label{fig: space time apl (a)}
     \end{subfigure}
     \hfill
     \begin{subfigure}[b]{1\textwidth}
         \centering
         \includegraphics[width=1\textwidth]{figures/trc_second_revision_new/heatmaps/waves_heatmap_adaptive_pl_400_40_seed_2_nudge_1.jpg}
         \caption{400~veh/km}
         \label{fig: space time apl (b)}
     \end{subfigure}
        \caption{Spatio-temporal speed plot for APL controllers with penetration of 20\% HDVs. The APL controllers reduce the intensity of traffic waves.}
        \label{fig: space time apl}
\end{figure}

The previous section also showed that the PL controller forms traffic waves with increasing HDV penetration. Therefore, it is also important to analyze the impact of the APL controller on this phenomenon. As shown in Figure~\ref{fig: space time apl}, the APL strategy does not completely remove the traffic waves but reduces their intensity. The long APL corridors in CM and NSCM help to reduce the traffic waves the most; however, the short APL corridors in FAM and SVAM are found to be less effective against traffic waves. At a very high vehicle density (400~$veh/km$), all APL strategies struggle to avoid forming traffic waves.

Overall, the results in this section show that for a relatively low number of HDVs (almost 40\% HDVs), an APL strategy that favors unhindered movements of groups of CAVs  like CM or NSCM produces better results. However, an APL strategy that favors the overtaking maneuvers of follower CAVs is more general and produces better results with higher HDV penetration. Nevertheless, even with the APL controller, the results indicate that penetration of at least 60\% CAVs into LFT is necessary before any major benefits of LFT to the overall traffic flow start to appear.

\begin{figure}[!tb]
     \centering
     \begin{subfigure}[b]{1\textwidth}
         \centering
         \includegraphics[width=\textwidth]{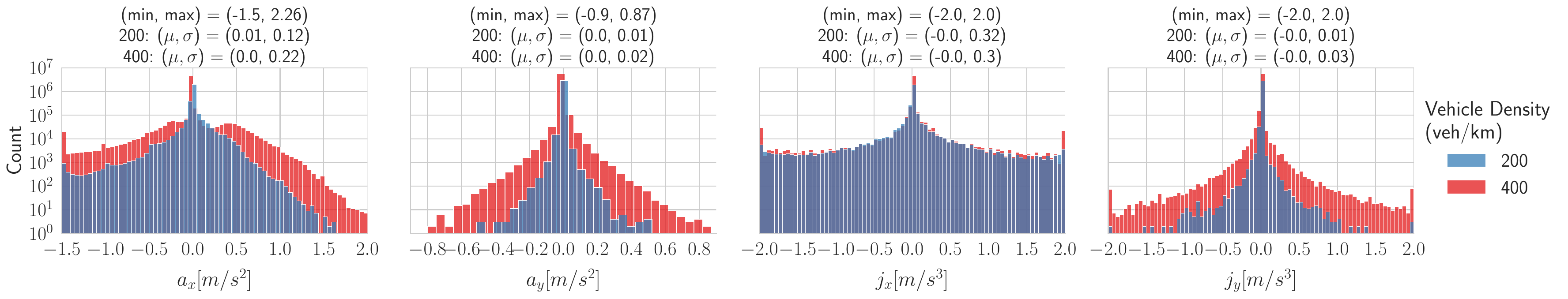}
         \caption{PL controller and all-CAVs scenarios.}
         \label{fig: comfort analysis (a)}
     \end{subfigure}
     \hfill
     \begin{subfigure}[b]{1\textwidth}
         \centering
         \includegraphics[width=\textwidth]{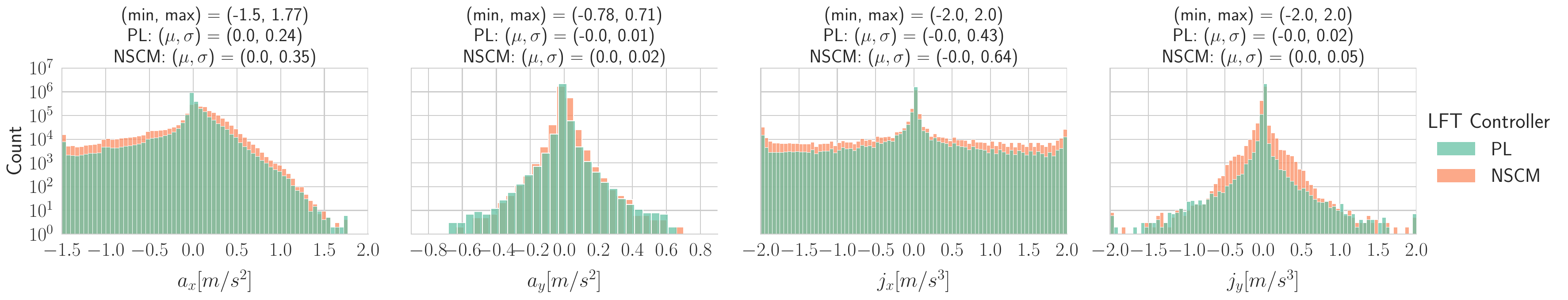}
         \caption{PL and NSCM controllers with 200~$veh/km$ and 20\% HDVs.}
         \label{fig: comfort analysis (b)}
     \end{subfigure}
    \hfill
     \begin{subfigure}[b]{1\textwidth}
         \centering
         \includegraphics[width=\textwidth]{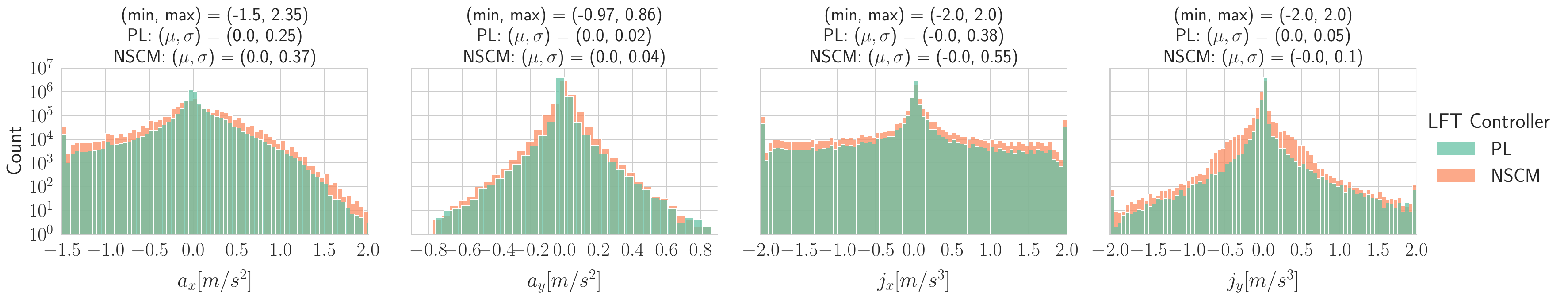}
         \caption{PL and NSCM controllers with 400~$veh/km$ and 20\% HDVs}
         \label{fig: comfort analysis (c)}
     \end{subfigure}
        \caption{Comparison of acceleration and jerk values for PL and NSCM controllers for all-CAVs and scenarios 20\% HDV penetration. The histogram analyzes the values of all CAVs recorded for all time steps. $j_x$ and $j_y$ represent the longitudinal and lateral jerk, respectively.}
        \label{fig: comfort analysis}
\end{figure}

% \usepackage{booktabs}
% \usepackage{multirow}
\begin{table}[]
\centering
\small
\begin{tabular}{@{}ccccllll@{}}
 &  &  & \multicolumn{5}{c}{\textbf{Method [CDF values in \%]}} \\ \midrule
\textbf{Vehicle Density (veh/km)} & \textbf{HDVs [\%]} & \textbf{TTC Threshold} & \textbf{PL} & \multicolumn{1}{c}{\textbf{CM}} & \multicolumn{1}{c}{\textbf{NSCM}} & \multicolumn{1}{c}{\textbf{FAM}} & \multicolumn{1}{c}{\textbf{SVAM}} \\ \midrule
\multirow{6}{*}{200} & \multirow{2}{*}{all-CAVs (0\%)} & 1.5~s & 0.001 & \multicolumn{1}{c}{} & \multicolumn{1}{c}{} & \multicolumn{1}{c}{} & \multicolumn{1}{c}{} \\
 &  & 3~s & 0.007 &  & \multicolumn{1}{c}{} & \multicolumn{1}{c}{} & \multicolumn{1}{c}{} \\ \cmidrule(l){2-8} 
 & \multirow{2}{*}{20\%} & 1.5~s & 0.008 & 0.013 & 0.013 & 0.016 & 0.017 \\
 &  & 3~s & 0.037 & 0.053 & 0.043 & 0.066 & 0.067 \\ \cmidrule(l){2-8} 
 & \multirow{2}{*}{50\%} & 1.5~s & 0.023 & 0.021 & 0.039 & 0.047 & 0.034 \\
 &  & 3~s & 0.074 & 0.076 & 0.136 & 0.131 & 0.119 \\ \midrule
\multirow{6}{*}{400} & \multirow{2}{*}{all-CAVs (0\%)} & 1.5~s & 0.120 & \multicolumn{1}{c}{} & \multicolumn{1}{c}{} & \multicolumn{1}{c}{} & \multicolumn{1}{c}{} \\
 &  & 3~s & 0.429 & \multicolumn{1}{c}{} & \multicolumn{1}{c}{} & \multicolumn{1}{c}{} & \multicolumn{1}{c}{} \\ \cmidrule(l){2-8} 
 & \multirow{2}{*}{20\%} & 1.5~s & 0.114 & 0.128 & 0.225 & 0.183 & 0.213 \\
 &  & 3~s & 0.409 & 0.451 & 0.792 & 0.638 & 0.738 \\ \cmidrule(l){2-8} 
 & \multirow{2}{*}{50\%} & 1.5~s & 0.103 & 0.103 & 0.120 & 0.168 & 0.167 \\
 &  & 3~s & 0.352 & 0.352 & 0.401 & 0.565 & 0.567 \\ \bottomrule
\end{tabular}
\caption{The values of the cumulative distribution function (CDF) for time-to-collision (TTC) thresholds. The CDF values are within reasonable ranges for safety. The negative TTC values were removed before calculating CDF.}
\label{tab: TTC values}
\end{table}

\subsection{Comfort and Safety}

The introduction of HDVs into the LFT environment and the APL controller increases lateral movements of CAVs. This behavior is visible in the vehicle trajectories in Figure \ref{fig: APL trajectory}. Since both comfort and safety are fundamental to traffic performance, it is necessary to evaluate whether these additional movements compromise either aspect.

Comfort is evaluated using vehicle acceleration and jerk, which should remain within accepted ranges to avoid discomfort to passengers. Figure \ref{fig: comfort analysis} compares acceleration and jerk for different control strategies and HDV penetration rates; the NSCM method is used as a representative APL controller since it provided the best performance and also led to high-frequency lateral movements in Figure~\ref{fig: APL trajectory}. In all-CAV scenarios, acceleration and jerk are smallest, with slightly wider spreads at higher vehicle densities. Introducing HDVs increases these spreads under the PL controller. For example, the standard deviation ($\sigma$) of longitudinal acceleration $\sigma(a_x)$ rises from 0.12~$m/s^2$ to 0.24~$m/s^2$, and the longitudinal jerk $\sigma(j_x)$ from 0.32~$m/s^3$ to 0.43~$m/s^3$ at 20\% HDV penetration and 200~$veh/km$. The same behavior is observed across both axes, vehicle densities, and penetration rates. This also indicates that the smooth movement of CAVs gets hindered by HDVs, causing them to accelerate and decelerate more often, leading to increased spread of acceleration and jerk.

The NSCM method produces an even higher spread of acceleration and jerk due to more frequent formation and reconfiguration of APL corridors, forcing CAVs to adjust positions more often. For example, at 20\% HDV and 400~$veh/km$, $\sigma(a_x)$ increases from 0.25~$m/s^2$ to 0.37~$m/s^2$ and $\sigma(j_x)$ from 0.38~$m/s^3$ to 0.55~$m/s^3$. Nonetheless, all values remain well within the hard limits used in the simulation for comfort. The $j_x$ may falsely appear to be uniformly distributed due to the log scale; however, $3\sigma$ limits indicate that 99\% of the $j_x$ are well below the jerk limit of (-2,2)~$m/s^3$.

While low jerk indicates smooth motion, it does not guarantee safe inter-vehicle spacing. Safety is therefore assessed using the time-to-collision (TTC) metric, defined as the gap between leader and follower divided by their relative speed. TTC represents the time until collision if both vehicles maintain their current speeds. Even though there are disagreements on the exact TTC thresholds for safe driving, values below 1.5~s are commonly classified as critical, and values below 3~s as cautionary \cite{national2023estimating}. Table \ref{tab: TTC values} shows the cumulative distribution function (CDF) values for TTC thresholds of 1.5~s and 3~s. In all-CAV scenarios, the proportion of leader–follower pairs with TTC less than 1.5~s is extremely low (0.001\% at 200~$veh/km$ and 0.12\% at 400~$veh/km$). Even with HDV penetration and different APL controllers, these percentages remain very low for both thresholds. The calculations also include the initialization period when vehicles are randomly positioned, which makes these results even more conservative.

Overall, the low TTC and jerk values indicate that the vehicles are driving at a sufficient distance from each other, and there was very limited sudden movement necessary to avoid collision. From which it can be concluded that the overall LFT setup used can potentially provide safe and comfortable driving.

\section{Summary}

The ever-increasing problem of traffic congestion in cities, coupled with advancements in AV technologies, has spurred the search for innovative solutions. Rather than investing in expensive and time-consuming new road infrastructures like beltways, there is a recent push to enhance the capacity and safety of existing infrastructure. One promising approach involves using CAVs. These vehicles can communicate with each other and the infrastructure to better coordinate their movements and improve traffic flow.

One significant way CAV technology can transform traffic management is through LFT. In LFT, vehicles coordinate their movements without relying on traditional fixed lanes. Recent studies have shown that LFT has the potential to significantly increase road capacity through the coordinated movements of CAVs. However, the transition to LFT may not be straightforward and could require the coexistence of independent traffic vehicles, either HDVs or simple AVs, during the transition phase.

This study examines the impact of HDVs on LFT using a microscopic simulation of a 1~$km$ ring road with a width of 10.2~meters. While HDVs make driving decisions based on individual benefits, CAVs use the PL controller for coordinated movements. The results indicate that a pure LFT scenario significantly improves road capacity compared to scenarios with only HDVs without lanes; a maximum flow of 16,700~$veh/h$ is observed for LFT compared to only 8,100~$veh/h$ for the latter. However, the flow improvement of LFT significantly drops as soon as HDVs are introduced into the system. Even a small penetration of HDVs, such as 5\%, reduces the maximum flow by 20\% (to 13,300 $veh/h$). The performance drop is significantly sharp for beginning 0-20\% HDVs, where at 20\% it drops by 40\%, and becomes half at 40\% HDV penetration. In literature, the LFT is also marked by its characteristic feature of avoiding the formation of traffic waves. However, this study found that with the penetration of HDVs, traffic waves start to appear in LFT and worsen with higher HDV penetrations and vehicle densities.

The study also introduced an APL controller approach to reduce the above performance drop. Unlike the simple PL controller, where the assigned PLs remain fixed, the APL controller adapts the PLs in the vicinity of HDVs. These areas with modified PLs are referred to as APL corridors. The study developed four APL variants that mainly differ on (1) the distance behind each HDV included in the APL corridor and (2) the conditions required for individual HDV whose vicinity is included in the corridor. The study found that the APL variants that favor the group movement of CAVs by forming long APL corridors perform better for a lower range of HDV penetration (up to almost 40\%). In comparison, the APL strategies that favor the overtaking maneuvers of individual CAVs perform better for higher HDV penetration. Overall, the NSCM strategy for APL controller provides the best performance. 

Even with the improved performance of the APL controller, the results indicate that the proportion of CAVs in LFT needs to be significantly high (almost 60\%) before the practical benefits of LFT with CAVs start to appear in terms of traffic flow.

\section{Limitations and Future Work}

Even though this study aimed to be as comprehensive as possible in its methodology, some limitations must be considered when interpreting the outcomes. First, the study used a specific model for the HDVs, which may not accurately represent real driving behavior without lanes, especially in LFT scenarios. A major limitation of the HDVs used is their complete inability to communicate with one another. In reality, HDVs may use signals such as honking or flashing headlights to indicate their intention to overtake the leading vehicle. Similarly, an HDV can also partially observe the movements of CAVs and adjust the driving behavior accordingly. However, the HDV model in this study does not consider such HDV-HDV or CAV-HDV interactions, which may not represent real-world behavior.

Additionally, the driving behavior of individual humans varies significantly, while the study model used similar parameters for all HDVs. The important parameters for the HDVs are the reaction time, minimum safety gap, and deceleration capability for calculating safe velocities. These can have a significant impact on outcomes. The same applies to CAVs, i.e., the study used a specific LFT controller and set of parameter values for the experiments. The outcomes may differ significantly if these parameters are changed. It is also possible that the impacts of HDVs are amplified due to the use of a ring road; even with a small proportion of HDVs, the CAVs repeatedly encounter the HDVs in each rotation, leading to an amplified effect on upstream vehicles.

The HDV model can be enhanced in the future to better represent human behavior in LFT scenarios. Driving simulator studies of LFT may be crucial for this purpose \cite{sekeranInvestigatingLaneFreeTraffic2023}, especially for estimating the reaction time and deceleration for HDVs in an LFT environment. The impact of HDVs should also be studied under different simulation scenarios to provide further insights into their effects. Similarly, the impact of HDVs on other LFT controllers should be investigated. It is important to note that the previously concluded penetration rate of 60\% CAVs to start seeing the major benefits of LFT indicates that LFT controller design should also consider boundary cases, such as handling situations when not all vehicles are CAVs. Even though APL controller reduces the impact of HDVs, there is still potential to further improve the LFT controller for these cases. The inclusion of HDVs makes safety considerations even more critical, as the LFT controller cannot directly manage these vehicles. Even though the current study considered acceleration and jerk limits for improved saftey and comfort, there is significant potential to improve them further. In this context, incorporating a safe acceleration formulation for the lateral axis—similar to that used for the longitudinal axis—could be beneficial. Additionally, the future work should also consider minimizing the lateral movements of APL strategies. The frequent activation and deactivation of APL corridors can reduce passenger comfort which can be looked into in the future for forming stable and long-standing APL corridors in-between HDVs for enhanced passenger comfort.

\section*{CRediT authorship contribution statement}

Arslan Ali Syed: Writing – review \& editing, Writing – original draft, Visualization, Validation, Software, Methodology. Majid Rostami-Shahrbabaki: Writing – review \& editing, Supervision, Conceptualization. Klaus Bogenberger: Writing – review, Supervision, Conceptualization, Funding acquisition.

\section*{Declaration of competing interest}
The authors declare that they have no known competing financial interests or personal relationships that could have appeared to influence the work reported in this paper.

\section*{Declaration of generative AI and AI-assisted technologies in the writing process}

During the preparation of this work, the authors used Grammarly and Microsoft Copilot to improve the language of the draft they prepared. After using this tool/service, the authors reviewed and edited the content as needed and take full responsibility for the content of the published article.

\section*{Acknowledgements}
This work is based on the project ”Simulation and organization of future lane-free traffic” funded by German research foundation (DFG), under the project number BO 5959/1-1.

\bibliography{literature.bib} 

\end{document}